\documentclass[acmsmall, nonacm]{acmart}

\acmJournal{PACMPL}
\startPage{1}

\setcopyright{none}

\usepackage{booktabs}   
\usepackage{subcaption} 
\usepackage{anyfontsize}                        
\usepackage{nameref}
\usepackage{pftools}
\usepackage{hyperref}
\usepackage{iris}
\usepackage{custom}
\usepackage{heaplang}
\usepackage{writing}
\usepackage{todonotes}
\usepackage[inline]{enumitem}
\usepackage{amsmath}
\usepackage{dsfont}
\usepackage{amsfonts}
\usepackage{amsthm}
\usepackage{stmaryrd}
\usepackage{mathpartir}
\usepackage{nameref}
\usepackage{wrapfig}
\usepackage{array}
\usepackage{mathrsfs}
\usepackage{xspace}
\usepackage{pict2e}
\usepackage{tikz}
\usepackage{cleveref}
\usepackage{fontawesome}
\usepackage{graphicx}
\usepackage{listings}
\usepackage[normalem]{ulem}
\usetikzlibrary{positioning, calc, matrix, arrows.meta, arrows}

\definecolor{backcolour}{rgb}{0.98,0.95,0.90}
\lstdefinestyle{mystyle}{
    language=[Objective]Caml,
    frame=single,
    framerule=0.5pt,
    alsoletter=?,
    backgroundcolor=\color{backcolour},
    showspaces=false,
    showstringspaces=false,
    showtabs=false,
    breakatwhitespace=false,
    breaklines=true,
    captionpos=b,
    keepspaces=true,
    columns=flexible,
    tabsize=2,
    basicstyle=\ttfamily\scriptsize,
    commentstyle=\color{green!60!black},
    escapeinside={(*@}{@*)},
    literate= {false}{{false}}5
              {true}{{true}}4
              {unique}{{\textbf{unique}}}6
              {aliased}{{\textbf{aliased}}}7
              {once}{{\textbf{once}}}4
              {many}{{\textbf{many}}}4
              {local}{{\textbf{local}}}5
              {global}{{\textbf{global}}}5
              {borrow}{{\textbf{borrow}}}6
              {ref}{{\textbf{ref}}}3
              {ret}{{\textbf{ret}}}3
              {effect}{{\textbf{effect}}}6
              {region}{{\textbf{region}}}6,
  numbers=left,
  stepnumber=1,
  numbersep=5pt,
  xleftmargin=1em,
  xrightmargin=0.6em,
}
\usepackage{caption}

\usepackage[belowskip=-12pt,aboveskip=4pt]{caption}

\allowdisplaybreaks

\begin{document}
\title{Yarrow: Reconciling Effect Handlers and Region-Based Memory Management}

\author{Anders Alnor Mathiasen}
\orcid{0009-0005-6587-5590}                
\affiliation{
  \institution{Aarhus University}            
  \country{Denmark}                          
}
\email{alnor@cs.au.dk}                  

\author{Amin Timany}
\orcid{0000-0002-2237-851X}             
\affiliation{            
  \institution{Aarhus University}
  \country{Denmark}                    
}
\email{timany@cs.au.dk}          

\author{Lars Birkedal}
\orcid{0000-0003-1320-0098}             
\affiliation{
  \institution{Aarhus University}            
  \country{Denmark}                    
}
\email{birkedal@cs.au.dk}          

\begin{abstract}
  We present a new ML-like programming language Yarrow with algebraic effects and region-based memory management. 
  Reconciling these programming language features into one language is challenging: the non-local control flow 
  of algebraic effects break the stack discipline of function calls and returns that
  region-based memory management relies on, and multi-shot effect handlers break 
  the invariant that regions can be exited at most once.
  We present a program logic, called Yarrow Logic (YL), 
  that supports safe and modular reasoning about regions in 
  the presence of one-shot and multi-shot effect handlers. 
  We prove the logic sound w.r.t. the operational semantics of
  Yarrow which is inspired by the runtime of OCaml but refined for regions.
  We use YL to prove correctness of a number of case studies with algebraic effects, 
  including checkpointing, asynchronous computation and a LIFO data structure implementation.
  Since all memory locations used in these case studies are allocated in regions, these case studies avoid using the less efficient garbage collected heap memory.
  We have formalized Yarrow's operational semantics, the Yarrow program logic, and all our case studies using the Iris separation logic framework on top of the Rocq Prover.
\end{abstract}



\maketitle

\makeatletter
\everydisplay\expandafter{\the\everydisplay \small}
\makeatother


\section{Introduction}
\label{sec:introduction}
In recent years, programming language designers have adopted algebraic effects and effect handlers
\citep{plotkin2009handlers, kammar2013handlers} in a wide variety of programming
languages, including OCaml \citep{sivaramakrishnan2021retrofitting}, Scala
\citep{brachthauser2020effekt}, WebAssembly \citep{phipps2023continuing}, Koka
\citep{leijen2014koka}, Links \citep{hillerstrom2016liberating}, Eff
\citep{bauer2015programming}, Frank \citep{convent2020doo} and Flix
\citep{flix, flix2}. Effect handlers give users the ability to suspend the computation by performing
effects, handle the effects using user-defined handlers, and resume the computation
through delimited continuations, where the effect handler is the delimiter,
either at most once (one-shot effects) or multiple times (multi-shot effects).
There are many successful applications of effect
handlers; \citet{dolan2015effective, dolan2017concurrent} show how effect
handlers can be used to implement a number of examples ranging from I/O
operations and web servers to asynchronous computation in OCaml. OCaml is known
for its efficient runtime implementation of effect handlers; delimited
continuations are implemented by pointers to stack segments called \emph{fibers}
\citep{sivaramakrishnan2021retrofitting}. Each effect handler creates a new
fiber when it is installed. This results in an efficient implementation of
one-shot effects as delimited continuations are captured by obtaining a pointer
to a stack segment without additional allocations; multi-shot effects
are less efficient and require copying of continuations.

\paragraph{Region-based memory management} Region-based memory management
is a memory allocation strategy used to avoid garbage collection pressure by
instead allocating and safely reclaiming memory in syntactically scoped program
fragments \citep{tofte1997region, tofte2004retrospective}. It is seen as a safer
way to avoid garbage collection overhead than full manual memory management, as
in C, because region-based memory management can be mostly automated
\citep{tofte2001programming}. Recently, OxCaml \citep{Georges25, Lorenzen24}
introduced region-based memory management into OCaml by using a rich type system based on
modes. One of these modes is the \emph{locality mode} that controls (when
applicable) whether a memory location is allocated using a \emph{local} allocation or a
\emph{global} allocation. Global allocations happen on the heap, while local
allocations are placed in memory regions that correspond to the scope of a
function.\footnote{OxCaml features special keywords provided whereby the programmer can have finer control over the allocation \citep{oxcaml-manual}.}
At runtime, \citet{birkedal1996region} showed that memory allocations of bounded regions can be 
placed on the call stack whereas unbounded regions can use linked-list structures on the heap.
The OxCaml extensions do not support
multi-shot effect handlers, and there does not exist any formalization of the interaction
between one-shot effect handlers and the OxCaml extensions.

\paragraph{Yarrow: Incorporating effect handlers with region-based memory management}
In this paper, we formalize an operational semantics for a new language, Yarrow,
an ML-like language with support for algebraic effects (one-shot and multi-shot)
based on fibers \citep{sivaramakrishnan2021retrofitting}, and with region-based
memory management for fine-grained memory control.
A first natural question is what the
combined semantics of effect handlers and region-based memory management should be. 
This is an inherently non-trivial question, as the non-local control 
flow that the delimited continuations of effect handlers introduce complicates reasoning about region configurations. Normally, regions are nested 
and follow a stack-like discipline, but with delimited continuations the control can jump between 
different groups of nested regions. Moreover, multi-shot effects break 
the invariant that regions can be exited at most once. 
For instance, consider the example below:
\begin{lstlisting}
region (                     (* create a new (*@\mdseries\color{green!60!black} region@*) *)
  let r = ref local 0 in     (* allocate a (*@\mdseries\color{green!60!black} reference@*) "r" pointing to "0" in this (*@\mdseries\color{green!60!black} region@*) *)
  do Foo r;                  (* perform the (*@\mdseries\color{green!60!black} effect@*) Foo with argument "r" *)
  !r)                        (* load from the (*@\mdseries\color{green!60!black} reference@*) "r" *)
\end{lstlisting}
This example uses a memory region created with the
$\langkw{region}$-construct; this means all local allocations within the
parentheses following $\langkw{region}$ happen inside the memory associated with
this newly created region. Thus, in this example $\ealloc{\localMode}{}$
allocates a reference inside the memory of the region which is automatically
freed when the scope of the region ends (this is contrary to
$\ealloc{\globalMode}{}$ which is a garbage-collected heap
allocation).\footnote{In \Cref{sec:background-regions}, we recap region-based
  memory management and the $\langkw{region}$-construct in more detail.}
Thereafter, the operation $\edo{\textlang{Foo}}{\textlang{r}}$ performs the
effect with name \textlang{Foo} and passes it \textlang{r} as an argument. Now, the
questions are: What happens when we load from the reference \textlang{r}? Is
it safe? Which value does the load produce? These are the types of questions we
answer in this paper; for this particular example, we argue that if
\textlang{Foo} is a one-shot effect, it is safe to load from \textlang{r} and
the return value is \textlang{0}, whereas if \textlang{Foo} were a multi-shot
effect, it would be unsafe to load from \textlang{r}. When \textlang{Foo}
is a one-shot effect, loading from \textlang{r} is safe because the region that holds \textlang{r} is captured as part of the continuation of \textlang{Foo}, and is restored when the continuation is invoked.
It is unsafe when \textlang{Foo} is a multi-shot effect because we want to support
region-based memory management where the memory of regions are freed upon exiting them; using \textlang{r} is thus unsafe because multi-shot continuations exit regions multiple
times, and therefore the first invocation of the continuation will free the region before all subsequent invocations.

\paragraph{Effect handlers in separation logic} The separation logic framework Iris 
\citep{irisjournal, iris, iris2, iris3} 
has been used to make informal claims about the behavior of programming languages 
precise, \eg for Rust \citep{RustBelt, dang2019rustbelt}, C \citep{mansky2024iris}, 
OCaml \citep{seassau2025formal, mevel2020cosmo, allain2026zoo}, and 
WebAssembly \citep{legoupil2024iris, rao2023iris}.
One successful application of Iris is to study effect handlers;
\citet{de2021separation} created a program logic for a language with effect handlers 
and mutable heap allocated references. 
\citet{de2021separation} enable modular proofs in that they establish, separately, correctness of 
effect handlers and of the programs producing the effects handled by those handlers.
This is achieved by specifying effect handlers through so called \emph{protocols}; this is similar to how a function's specification serves as a contract between the caller and the callee.
\paragraph{The Yarrow logic: Reasoning about effect handlers and regions}
In this paper we present a new program logic, the Yarrow Logic (YL), which supports modular reasoning about effect handlers and
region-based memory management in Yarrow. We use YL to specify and 
prove correctness of several examples featuring effect handlers, in order to exercise the Yarrow semantics, and a number of case studies, 
to show that effect handlers can safely avoid garbage collection by using region-based memory management.
The case studies include checkpointing, asynchronous computation and a LIFO data structure implementation.\footnote{We use the word LIFO here in place of \emph{stack} to avoid confusing with the (call) stack of the program that we will regularly mention in the paper.}
Reasoning about the language features of Yarrow is highly non-trivial mainly due to the following two primary challenges introduced by the non-well-bracketed control-flow of Yarrow, which YL solves 
\emph{while retaining the usual modularity of higher-order impredicative separation logic}:
\begin{description}[leftmargin=0pt, labelindent=0pt]
\item[Fine-grained reasoning about region-allocated memory] In region-based
  memory management we need to track the resources corresponding
  region-allocated memory and revoke them when the region goes out of scope.
  This tracking and revocation is very natural \cite{Georges25} when control
  flow is well-bracketed \citep{timany2024flow}. However, when control-flow is
  non-well-bracketed, \eg{} in the presence of effect handlers, tracking
  resources that should be reclaimed becomes very intricate, so much so that one
  may expect that the only way to cleanly track these resources is to forgo
  tracking individual memory locations that facilitates modular proofs, and instead explicitly reason about regions 
  using a single high-level logical unit with all information about regions.
  In YL, however, we show how
  reasoning about regions can be decomposed into (1) a logical unit representing configuration of fibers with associated regions,
  which only tracks the high-level changes to regions at revocation points, 
  and (2) stack-points-to propositions (similar
  to separation logic's points-to propositions used for tracking heap
  allocations), which allow for fine-grained (per memory location) reasoning about
  stack-allocated references.
  At the surface level YL's approach to enable
  fine-grained reasoning about region-allocated memory is not unlike existing
  approaches \citep{Georges25,Timany2018ST,Jespersen2017Effects}; what is
  particularly challenging here, and novel, is doing so in the presence of
  non-well-bracketed control-flow.
\item[The highly dynamic nature of fibers] The high-level configuration of
  fibers with regions that we track in YL to enable revocation is subject to sophisticated
  changes due to the way fibers can change upon suspension and resumption of
  computations in effect handlers; an effect handler may start and end several
  regions, or install and uninstall other effect handlers, before it uses its
  continuation. To solve this challenge in YL, we use the insight 
  that the configuration of fibers and regions should be a part of
  the contract between an effect handler and the program performing the effect
  --- continuations can capture regions when performing effects, 
  making this a point of temporary revocation for one-shot effects and permanent revocation 
  for multi-shot effects. 
  This solution has led us to redesign the
  protocol concept of \citet{de2021separation} to take into account the
  configuration of fibers and regions.
\end{description}

\paragraph{Contributions} In summary, the contributions of this paper are:
\begin{itemize}
\item An operational semantics for Yarrow, an ML-like language that supports region-based memory
  management in the presence of (one-shot and multi-shot) effect handlers (\Cref{sec:model}).
  The presence of effect handlers makes supporting region-based memory management intricate.
  Prior work on region-based memory management was designed with the stack discipline of calls and returns which is violated by the delimited continuations of effect handlers.
  Capturing all the nuances of these intricacies requires our operational semantics to be more fine-grained than was previously needed for modeling languages with either region-based memory management, or effect handlers.
\item A program logic Yarrow Logic (\Cref{sec:logic}), abbreviated YL,
  for proving correctness of Yarrow programs. Reasoning about Yarrow 
  is highly non-trivial as effect handlers break the property of 
  well-bracketed control flow, which in the past was relied upon for revocation 
  of resources tied to regions. 
  Despite the challenges that come with the new combination of language features, 
  YL retains the modularity and simplicity of existing program logics.
\item Specification and verification of a number of case studies, including a LIFO data structure
  implementation (\Cref{sec:stack},) checkpointing (\Cref{sec:checkpoint}), and
  asynchronous computation (\Cref{sec:async}).
  These case studies show that in certain applications programs written using effect handlers in Yarrow can replace garbage collected heap references with region allocated references.
  \item A mechanization of all results in this paper using the Rocq Prover. The Rocq mechanization accompanies the paper.
\end{itemize}
In \Cref{sec:background}, we summarize existing work on effect handler fibers 
(\Cref{sec:background-fibers}) and region-based memory management (\Cref{sec:background-regions}).


\section{Background}
\label{sec:background}

We proceed by explaining how effect handler fibers and region-based memory management work separately. 
For the presentation of examples and case studies, we use an OCaml like syntax, 
the calculus of our formalization is shown in \Cref{sec:semantics}.

\subsection{Background: Effect Handler Fibers}
\label{sec:background-fibers}
In the runtime of OCaml \citep{sivaramakrishnan2021retrofitting}, 
the stack is made up of a list of segments called \emph{fibers}. 
A new fiber is created and appended to the list of fibers
when an effect handler is installed. 
We proceed to explain effect fibers using the \textlang{Choose} and \textlang{State} effects: 
the \textlang{Choose} effect is used for backtracking in programs, and the \textlang{State} effect 
gives users access to a piece of state with the implementation hidden behind the effect abstraction.
\paragraph{Choose and State effects.} 
Consider the effect handlers in \Cref{fig:choose-state} for the \textlang{Choose} and \textlang{State} effects
and ignore the code in the comments for now (we consider this code in \Cref{sec:model}).
\begin{figure}[h]
\vspace{-3mm}
\begin{lstlisting}
let handle_choose f = (* region ( *)
  try (* (global, many) *) f () with 
  | effect Choose p k -> let x = k p.1 in let y = k p.2 in (x, y) | ret x -> x (* ) *)
  
let handle_state init f = (* region ( *)
  let r = ref (* local *) init in
  try (global, once) f () with
  | effect State arg k -> (match arg with | Read -> k !r | Write x -> (r <- x; k ()))
  | ret x -> x (* ) *)

let example1 () = (* region ( *)
  let x = do Choose (1, 2) in do State (Write ((do State Read) + x)); do State Read (* ) *)

let handle_example1 () = (* region ( *)
  assert (handle_state 0 (fun () => handle_choose example1) = (1, 3)) (* ) *)
\end{lstlisting}
\caption{\textlang{Choose} and \textlang{State} effect handlers with a closed example program.}
\label{fig:choose-state}
\end{figure}
In the code, the \textlang{handle\_choose} definition handles the \textlang{Choose} effect 
using the \langkw{try}-construct on line 2. Using the \langkw{try}-construct, it is safe to perform the \textlang{Choose} effect (on line 12)
inside the function for which we install the effect handler (on line 15). 
Much like a \langkw{match} case, the \langkw{try}-construct has two cases: one for when \textlang{f} performs the \textlang{Choose} effect 
(the handler branch), and one for when \textlang{f} terminates with a value (the return branch). 
The handler branch takes as arguments the user-provided argument \textlang{p}, and a delimited continuation \textlang{k}, 
where the \langkw{try}-construct acts as the delimiter, for resuming computation inside \text{f} at the exact point where the effect was performed.
The effect \langkw{Choose} is a multi-shot effect meaning the continuation can be used more than once; 
the implementation in \Cref{fig:choose-state} makes use of this by evaluating the continuation \textlang{k} on both 
components of the argument pair \textlang{p}, after which it returns the result in a pair. 
The \textlang{State} effect is implemented on line 7 in \textlang{handle\_state} by internally using the heap allocated reference 
\textlang{r}. We use \textlang{Read} and \textlang{Write} as notation for sum types, to 
branch on whether the user of the effect wants to read or write to the state, 
\ie \textlang{Read} is \textlang{InjL ()} and \textlang{Write x} is \textlang{InjR x}.
\paragraph{Effect fibers} 
Let us consider the layout of fibers at different points in the execution of the closed example on line 14 in 
\Cref{fig:choose-state}. The \textlang{example1} definition on line 11 uses effect handlers for \textlang{Choose}
and \textlang{State}. On line 14, \textlang{handle\_example1} executes \textlang{example1} using the two effect handlers 
\textlang{handle\_choose} and \textlang{handle\_state}. The state is initialized to \textlang{0}
which results in the computation returning the pair \textlang{(1, 3)}, as both executions 
created on line 12 with the \textlang{Choose} effect share the same state of the \textlang{State} effect which behind the effect abstraction 
is a heap allocated reference. Just before line 12 is executed 
the fiber configuration looks like this:
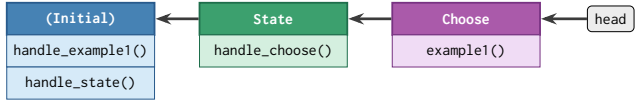
\begin{center}
  \scalebox{0.85}{\begin{tikzpicture}[
  header/.style = {
    draw, thick,
    minimum width  = 2.3cm,
    minimum height = 0.5cm,
    inner sep      = 4pt,
    text centered,
    font = \scriptsize\bfseries,
    text = white
  },
  inner/.style = {
    draw,
    minimum width  = 2.3cm,
    minimum height = 0.5cm,
    inner sep      = 3pt,
    text centered,
    align = center,
    font  = \scriptsize
  },
  llarrow/.style = {
    -{Stealth[length=7pt, width=5pt]},
    line width = 1.2pt,
    color = arrowC
  }
]
 
\def\Ax{0}
\node[header, fill=headerA, draw=borderA] (Ah) at (\Ax,  0.000) {\textlang{(Initial)}};
\node[inner,  fill=innerA,  draw=borderA] (A1) at (\Ax, -0.5)
  {\textlang{handle\_example1()}};
\node[inner,  fill=innerA,  draw=borderA] (A2) at (\Ax, -1.0)
  {\textlang{handle\_state()}};
 
\def\Bx{3.0}
\node[header, fill=headerB, draw=borderB] (Bh) at (\Bx,  0.000) {\textlang{State}};
\node[inner,  fill=innerB,  draw=borderB] (B1) at (\Bx, -0.5)
  {\textlang{handle\_choose()}};

\def\Cx{6.0}
\node[header, fill=headerC, draw=borderC] (Ch) at (\Cx,  0.000) {\textlang{Choose}};
\node[inner,  fill=innerC,  draw=borderC] (C1) at (\Cx, -0.5)
  {\textlang{example1()}};
 
\draw[llarrow] (Bh.west) -- (Ah.east);
\draw[llarrow] (Ch.west) -- (Bh.east);
\node[draw, rounded corners=2pt, fill=gray!15, font=\scriptsize\ttfamily, minimum height=0.4cm]
  (head) at (\Cx + 2.25, 0) {head};
\draw[llarrow] (head.west) -- (Ch.east);
  
\end{tikzpicture}}
\end{center}
The layout of the fibers naturally follows the order in which they were installed 
(\ie they reflect the order in \textlang{handle\_example1}): 
the head of the list of fibers always points to the most recently installed fiber, in this case 
the fiber for the \textlang{Choose} effect. Inside the fiber installed by the \textlang{Choose} effect handler
in \textlang{handle\_choose}, we have the call stack for the \textlang{example1} function. 
The \textlang{Choose} fiber then points back to the fiber installed by the \textlang{State} effect handler in 
\textlang{handle\_state}. This fiber contains the call stack for the \textlang{handle\_choose} function.
Lastly, the \textlang{State} fiber points back to the \emph{initial} fiber; there is always an initial 
fiber that contains the call stack of the top most functions of programs before any effect handlers are
installed. In this case, the initial fiber contains the call stack for the \textlang{handle\_state}
and \textlang{handle\_example1} functions.

When the \textlang{Choose} effect is performed on line 12 in \Cref{fig:choose-state}, 
computation is suspended and the control is transferred to the effect handler inside \textlang{handle\_choose}. 
This has the following effect on the fiber configuration:
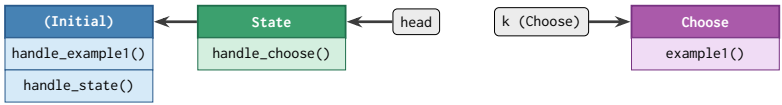
\begin{center}
  \scalebox{0.85}{\begin{tikzpicture}[
  header/.style = {
    draw, thick,
    minimum width  = 2.3cm,
    minimum height = 0.5cm,
    inner sep      = 4pt,
    text centered,
    font = \scriptsize\bfseries,
    text = white
  },
  inner/.style = {
    draw,
    minimum width  = 2.3cm,
    minimum height = 0.5cm,
    inner sep      = 3pt,
    text centered,
    align = center,
    font  = \scriptsize
  },
  llarrow/.style = {
    -{Stealth[length=7pt, width=5pt]},
    line width = 1.2pt,
    color = arrowC
  }
]
 
\def\Ax{0}
\node[header, fill=headerA, draw=borderA] (Ah) at (\Ax,  0.000) {\textlang{(Initial)}};
\node[inner,  fill=innerA,  draw=borderA] (A1) at (\Ax, -0.5)
  {\textlang{handle\_example1()}};
\node[inner,  fill=innerA,  draw=borderA] (A2) at (\Ax, -1.0)
  {\textlang{handle\_state()}};
 
\def\Bx{3.0}
\node[header, fill=headerB, draw=borderB] (Bh) at (\Bx,  0.000) {\textlang{State}};
\node[inner,  fill=innerB,  draw=borderB] (B1) at (\Bx, -0.5)
  {\textlang{handle\_choose()}};

\def\Cx{9.75}
\node[header, fill=headerC, draw=borderC] (Ch) at (\Cx,  0.000) {\textlang{Choose}};
\node[inner,  fill=innerC,  draw=borderC] (C1) at (\Cx, -0.5)
  {\textlang{example1()}};
 
\draw[llarrow] (Bh.west) -- (Ah.east);
\node[draw, rounded corners=2pt, fill=gray!15, font=\scriptsize\ttfamily, minimum height=0.4cm]
  (head) at (\Bx + 2.25, 0) {head};
\draw[llarrow] (head.west) -- (Bh.east);
\node[draw, rounded corners=2pt, fill=gray!15, font=\scriptsize\ttfamily, minimum height=0.4cm]
  (cont) at (\Cx - 2.55, 0) {\textlang{k (Choose)}};
\draw[llarrow] (cont.east) -- (Ch.west);
  
\end{tikzpicture}}
\end{center}
Above, we see that the head of the list now points to the \textlang{State} fiber. 
The delimited continuation \textlang{k} used inside the effect handler \textlang{Choose} is a pointer to the 
\textlang{Choose} effect fiber; when an effect is performed, we traverse the list of fibers from the head 
until the correct fiber is found and create the continuation as a pointer to this fiber. 
The continuation \textlang{k} is thus a pointer to all the call stacks required to resume computation 
at the point where it was suspended when the \textlang{Choose} effect was performed. 
Similarly, when the \textlang{State} effect is performed, \eg 
on line 12 in \Cref{fig:choose-state}, the fiber configuration looks like this: 
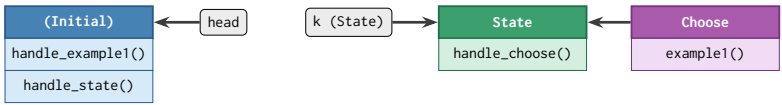
\begin{center}
  \scalebox{0.85}{\begin{tikzpicture}[
  header/.style = {
    draw, thick,
    minimum width  = 2.3cm,
    minimum height = 0.5cm,
    inner sep      = 4pt,
    text centered,
    font = \scriptsize\bfseries,
    text = white
  },
  inner/.style = {
    draw,
    minimum width  = 2.3cm,
    minimum height = 0.5cm,
    inner sep      = 3pt,
    text centered,
    align = center,
    font  = \scriptsize
  },
  llarrow/.style = {
    -{Stealth[length=7pt, width=5pt]},
    line width = 1.2pt,
    color = arrowC
  }
]
 
\def\Ax{0}
\node[header, fill=headerA, draw=borderA] (Ah) at (\Ax,  0.000) {\textlang{(Initial)}};
\node[inner,  fill=innerA,  draw=borderA] (A1) at (\Ax, -0.5)
  {\textlang{handle\_example1()}};
\node[inner,  fill=innerA,  draw=borderA] (A2) at (\Ax, -1.0)
  {\textlang{handle\_state()}};
 
\def\Bx{6.75}
\node[header, fill=headerB, draw=borderB] (Bh) at (\Bx,  0.000) {\textlang{State}};
\node[inner,  fill=innerB,  draw=borderB] (B1) at (\Bx, -0.5)
  {\textlang{handle\_choose()}};

\def\Cx{9.75}
\node[header, fill=headerC, draw=borderC] (Ch) at (\Cx,  0.000) {\textlang{Choose}};
\node[inner,  fill=innerC,  draw=borderC] (C1) at (\Cx, -0.5)
  {\textlang{example1()}};
 
\draw[llarrow] (Ch.west) -- (Bh.east);
\node[draw, rounded corners=2pt, fill=gray!15, font=\scriptsize\ttfamily, minimum height=0.4cm]
  (head) at (\Ax + 2.25, 0) {head};
\draw[llarrow] (head.west) -- (Ah.east);
\node[draw, rounded corners=2pt, fill=gray!15, font=\scriptsize\ttfamily, minimum height=0.4cm]
  (cont) at (\Bx - 2.55, 0) {\textlang{k (State)}};
\draw[llarrow] (cont.east) -- (Bh.west);
  
\end{tikzpicture}}
\end{center}
When the computation is resumed inside the effect handlers of 
\textlang{State} or \textlang{Choose} by using the continuation, the fibers that the continuation points to 
are reattached to the list of fibers that the head points to. 
Because the \textlang{Choose} effect is a multi-shot effect, in that its continuation \textlang{k} is 
used many times in the effect handler on line 3 in \Cref{fig:choose-state}, 
we must make copies of the fibers that the continuation of the \textlang{Choose} effect
points to. Thus, when the continuation is used to resume computation in the 
\textlang{Choose} effect, we create a copy of the fibers that the continuation
points to and attach those to the list of fibers that the head points to which 
still leaves the original fibers in memory:
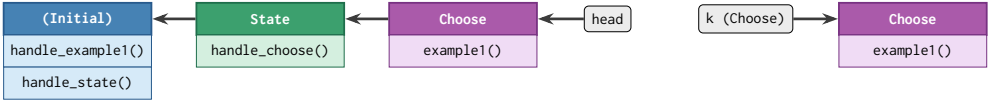
\begin{center}
  \scalebox{0.85}{\begin{tikzpicture}[
  header/.style = {
    draw, thick,
    minimum width  = 2.3cm,
    minimum height = 0.5cm,
    inner sep      = 4pt,
    text centered,
    font = \scriptsize\bfseries,
    text = white
  },
  inner/.style = {
    draw,
    minimum width  = 2.3cm,
    minimum height = 0.5cm,
    inner sep      = 3pt,
    text centered,
    align = center,
    font  = \scriptsize
  },
  llarrow/.style = {
    -{Stealth[length=7pt, width=5pt]},
    line width = 1.2pt,
    color = arrowC
  }
]
 
\def\Ax{0}
\node[header, fill=headerA, draw=borderA] (Ah) at (\Ax,  0.000) {\textlang{(Initial)}};
\node[inner,  fill=innerA,  draw=borderA] (A1) at (\Ax, -0.5)
  {\textlang{handle\_example1()}};
\node[inner,  fill=innerA,  draw=borderA] (A2) at (\Ax, -1.0)
  {\textlang{handle\_state()}};
 
\def\Bx{3.0}
\node[header, fill=headerB, draw=borderB] (Bh) at (\Bx,  0.000) {\textlang{State}};
\node[inner,  fill=innerB,  draw=borderB] (B1) at (\Bx, -0.5)
  {\textlang{handle\_choose()}};

\def\Cx{6.0}
\node[header, fill=headerC, draw=borderC] (Ch) at (\Cx,  0.000) {\textlang{Choose}};
\node[inner,  fill=innerC,  draw=borderC] (C1) at (\Cx, -0.5)
  {\textlang{example1()}};

\def\Dx{13.00}
\node[header, fill=headerC, draw=borderC] (Dh) at (\Dx,  0.000) {\textlang{Choose}};
\node[inner,  fill=innerC,  draw=borderC] (D1) at (\Dx, -0.5)
  {\textlang{example1()}};

\draw[llarrow] (Bh.west) -- (Ah.east);
\draw[llarrow] (Ch.west) -- (Bh.east);
\node[draw, rounded corners=2pt, fill=gray!15, font=\scriptsize\ttfamily, minimum height=0.4cm]
  (head) at (\Cx + 2.25, 0) {head};
\draw[llarrow] (head.west) -- (Ch.east);
\node[draw, rounded corners=2pt, fill=gray!15, font=\scriptsize\ttfamily, minimum height=0.4cm]
  (cont) at (\Dx - 2.6, 0) {\textlang{k (Choose)}};
\draw[llarrow] (cont.east) -- (Dh.west);
  
\end{tikzpicture}}
\end{center}
\subsection{Background: Region-based Memory Management}
\label{sec:background-regions}
In this subsection, we focus on regions (without effects).
In the introduction (\Cref{sec:introduction}), we already saw a glimpse of the kind of regions 
we consider in this paper, namely regions {\`a} la \citet{tofte1997region} but with a  
construct similar to that of \citet{Lorenzen24} and \citet{Georges25}. 
The work by \citet{Lorenzen24} and \citet{Georges25} differentiates between two types of allocations: 
local allocations in regions and global allocations on the heap.
To be precise, the construct $\eregion{(...)}$, as seen in the introduction (\Cref{sec:introduction}), 
is used to create lexically scoped regions with a 
stack-like discipline. All local allocations are associated with the lexically nearest enclosing region construct.
A \emph{locality mode} $\localVar$, which can be either \textlang{\localMode} or \textlang{\globalMode}, 
is used to differentiate between local allocations in regions and global heap allocations; 
a reference to a value \textlang{\val} is allocated using \langkw{ref} \textlang{\localVar ~ \val} where 
\textlang{\localVar} is the locality mode.
We remark that the mode notation used here are arguments to program constructs and not types 
(our examples are untyped).\footnote{In the work by \citet{Georges25} they use mode annotations on types with the \langkw{@} character
and superscript notation for modes of program constructs.}
To explain the semantics of regions, and in particular nested regions, 
we consider the example in \Cref{fig:region-simpl}.
\begin{figure}[h]
\vspace{-2mm}
\begin{lstlisting}
let foo x = region (let y = ref local 1 in (x, y))

let bar () = region (let x = ref local 0 in let p = foo x in !p.1)
\end{lstlisting}
\caption{Nested Regions Example.}
\label{fig:region-simpl}
\end{figure}
In this example, both functions create a region around their function bodies. 
Indeed, as a default we associate regions with function bodies
(insertion of \langkw{region} can be delegated to the compiler), but exceptions can be made to this to principle, 
we will see one example of where such an exception can be useful later in \Cref{sec:async}.
The function \textlang{foo} returns a pair; the first entry in the pair is the argument that \textlang{foo}
receives, and the second entry is a reference, locally allocated in its own region. 
The function \textlang{bar} creates a local allocation \textlang{x} in its region, and
then calls \textlang{foo} with \textlang{x} before it loads from the first value of the pair 
that it has gotten from the call to  \textlang{foo}. 
Below, we visualize how the stack evolves throughout the execution of the function \textlang{bar}, 
the arrows signify changes to the stack configuration:
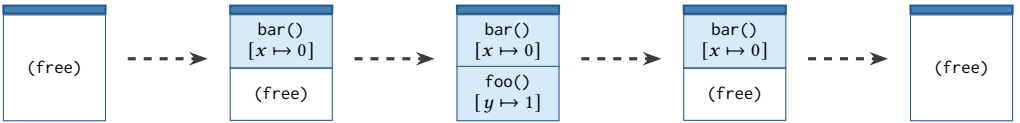
\begin{center}
  \begin{tikzpicture}[
  header/.style = {
    draw, thick,
    minimum width  = 1.35cm,
    minimum height = 3pt,
    inner sep      = 0pt,
    text centered,
    font = \scriptsize\bfseries,
    text = white
  },
  inner/.style = {
    draw,
    minimum width  = 1.35cm,
    minimum height = 0.65cm,
    inner sep      = 3pt,
    text centered,
    align = center,
    font  = \scriptsize
  },
  llarrow/.style = {
    -{Stealth[length=7pt, width=5pt]},
    line width = 1.2pt,
    color = arrowC, 
    dashed, 
    shorten >=8pt,
    shorten <=8pt
  },
]
 
\def\Ax{-6}
\node[header, fill=headerA, draw=borderA] (Ah) at (\Ax,  0.000) {};
\node[draw, font=\scriptsize, align=center, minimum width  = 1.35cm,
  minimum height = 1.4cm, draw=borderA] (A1) at (\Ax,  -0.77) {\textlang{(free)}};

\def\Bx{-3}
\node[header, fill=headerA, draw=borderA] (Bh) at (\Bx,  0.000) {};
\node[inner,  fill=innerA,  draw=borderA] (B1) at (\Bx, -0.42)
   {\textlang{bar()} \\ $[x \stackptr 0]$};
\node[inner,  draw=borderA, minimum height = 0.69cm] (B2) at (\Bx, -1.12)
  {\textlang{(free)}};

\def\Cx{0}
\node[header, fill=headerA, draw=borderA] (Ch) at (\Cx,  0.000) {};
\node[inner,  fill=innerA,  draw=borderA] (C1) at (\Cx, -0.42)
  {\textlang{bar()} \\ $[x \stackptr 0]$};
\node[inner,  fill=innerA,  draw=borderA] (C2) at (\Cx, -1.11)
  {\textlang{foo()} \\ $[y \stackptr 1]$};

\def\Dx{3}
\node[header, fill=headerA, draw=borderA] (Dh) at (\Dx,  0.000) {};
\node[inner,  fill=innerA,  draw=borderA] (D1) at (\Dx, -0.42)
  {\textlang{bar()} \\ $[x \stackptr 0]$};
\node[inner,  draw=borderA, minimum height = 0.69cm] (D3) at (\Dx, -1.12)
  {\textlang{(free)}};

\def\Ex{6}
\node[header, fill=headerA, draw=borderA] (Eh) at (\Ex,  0.000) {};
\node[draw, font=\scriptsize, align=center, minimum width  = 1.35cm,
  minimum height = 1.4cm, draw=borderA] (E1) at (\Ex,  -0.77) {\textlang {(free)}};

\draw[llarrow] ([yshift=-0.65cm]Ah.east) -- ([yshift=-0.65cm]Bh.west);
\draw[llarrow] ([yshift=-0.65cm]Bh.east) -- ([yshift=-0.65cm]Ch.west);
\draw[llarrow] ([yshift=-0.65cm]Ch.east) -- ([yshift=-0.65cm]Dh.west);
\draw[llarrow] ([yshift=-0.65cm]Dh.east) -- ([yshift=-0.65cm]Eh.west);

\end{tikzpicture}
\end{center}
Each stack configuration consists of zero or more stack frames.
Under each stack frame, we also write the allocations made in the region associated with the function that uses this stack frame.
We emphasize that regions are not necessarily a part of the stack frame, the memory underlying regions can be placed elsewhere on the heap
(usually bounded regions go on the call stack and unbounded on the heap, see \citet{birkedal1996region}).
However, the regions follow a stack discipline, enforced by the  \langkw{region}-construct:
before \textlang{foo} returns the pair in \Cref{fig:region-simpl}, the stack configuration is as the third configuration 
above with the call stack, and thus the regions of both functions are still in memory. After \textlang{foo} returns its pair and execution 
resumes in \textlang{bar}, we move along to the fourth stack configuration, where the region of \textlang{foo} has been freed, and 
its contents deallocated. When \textlang{bar} loads from the first entry in the pair it received from \textlang{foo}, 
it is safe as the first entry is \textlang{x}, which is still alive, in \textlang{bar}'s own region.
Note that if the load had been from the second entry in the pair, it would have been unsafe, 
as \textlang{y} lived in the region of \textlang{foo}, which was freed when \textlang{foo} returned.


\section{Operational Semantics}
\label{sec:model}
In this section, we explain our new operational semantics for Yarrow,
an ML-like language with effect handlers and region-based memory management.
Before we explain the details of the operational semantics in (\Cref{sec:semantics}), 
we give an overview of the ideas behind the operational semantics in (\Cref{sec:overview}).
\subsection{Overview}
\label{sec:overview}
We introduce our new operational semantics using two examples, 
the first example focuses on one-shot effects and the second example focuses 
on multi-shot effects.
\paragraph{One-shot effects}
For one-shot effects, we illustrate the point that continuations capture regions,
and we can resume the computation with captured regions as if nothing happened to these regions
(this is not the case when we consider multi-shot effects next). 
Consider the \emph{Increment First Example} in \Cref{fig:inc-fst-impl}.
\begin{figure}[h]
\vspace{-2.5mm}
\begin{lstlisting}
let handle_inc_fst f = region ( 
  let x = ref local 0 in
  try (global, once) f x with | effect IncFst p k -> (p.1 <- !p.1 + 1); k () | ret x -> x)  
  
let example2 x = region (
  let y = ref local 1 in let p = (x, y) in do IncFst p; assert (!p.1 + !p.2 = 2))

let handle_example2 () = region (handle_inc_fst example2)
\end{lstlisting}
\caption{Increment First Example.}
\label{fig:inc-fst-impl}
\vspace{-1mm}
\end{figure}
The figure contains two definitions \textlang{handle\_inc\_fst} and \textlang{example2};
at the end of the figure, \textlang{handle\_inc\_fst} is applied to \textlang{example2} 
in \textlang{handle\_example2}. 
After creating a region, \textlang{handle\_inc\_fst} allocates a reference \textlang{x} pointing to \textlang{0}
locally in this region. Next, an effect handler with operation \textlang{IncFst} is installed around 
the application of the function argument to the newly allocated reference \textlang{x}. 
In our ML-like language, we use mode notation with the \langkw{try}-construct: 
the \langkw{try}-construct takes a pair of modes where the first entry is a locality mode that 
can be \textlang{\globalMode} or \textlang{\localMode}, just like with references, 
and the second entry is an affinity mode that can be either \textlang{\onceMode} or \textlang{\manyMode}.
The locality mode controls whether the closure of the continuation \textlang{k} in the handler branch 
is allocated locally in the region where the effect handler is situated, or globally on the heap. 
Later in \Cref{sec:semantics}, we will also see how function closures can be allocated in regions 
instead of on the heap, but all the functions we have seen so far have non-empty closures.  
The affinity mode controls whether the effect is a one-shot or multi-shot effect, 
\ie how many times the continuation \textlang{k} can be called in the handler branch.
For the effect handler in the \textlang{handle\_inc\_fst} definition of \Cref{fig:inc-fst-impl}, 
we use the \textlang{(\globalMode, \onceMode)} combination; \textlang{\onceMode} because 
it is a one-shot effect, and for simplicity to focus on the core parts of regions and effects, 
we use the \textlang{\globalMode} mode in this example
(our case studies use continuations with locally allocated closures).
The handler branch for the \textlang{IncFst} effect handler takes a pair and increments 
a reference in the pair's first entry. No assumptions are made about the second entry in the pair.
Moving our attention to the \textlang{example2} definition, this function
performs the effect \textlang{IncFst} with argument \textlang{p}. 
The first entry of the pair \textlang{p} is the function argument \textlang{x} of \textlang{example2}, 
whereas the second entry is a reference \textlang{y} locally allocated inside the region \textlang{example2}.
The configuration of fibers just before $\edo{\textlang{IncFst}}{\textlang{p}}$ 
is executed is this:
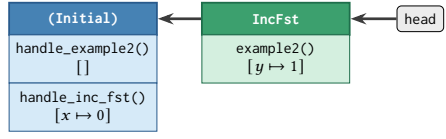
\begin{center}
  \scalebox{0.85}{\begin{tikzpicture}[
  header/.style = {
    draw, thick,
    minimum width  = 2.3cm,
    minimum height = 0.5cm,
    inner sep      = 4pt,
    text centered,
    font = \scriptsize\bfseries,
    text = white
  },
  inner/.style = {
    draw,
    minimum width  = 2.3cm,
    minimum height = 0.75cm,
    inner sep      = 3pt,
    text centered,
    align = center,
    font  = \scriptsize
  },
  llarrow/.style = {
    -{Stealth[length=7pt, width=5pt]},
    line width = 1.2pt,
    color = arrowC
  }
]
 
\def\Ax{0}
\node[header, fill=headerA, draw=borderA] (Ah) at (\Ax,  0.000) {\textlang{(Initial)}};
\node[inner,  fill=innerA,  draw=borderA] (A1) at (\Ax, -0.625)
  {\textlang{handle\_example2()} \\ $[]$};
\node[inner,  fill=innerA,  draw=borderA] (A2) at (\Ax, -1.375)
  {\textlang{handle\_inc\_fst()} \\ $[x \stackptr 0]$};
 
\def\Bx{3.0}
\node[header, fill=headerB, draw=borderB] (Bh) at (\Bx,  0.000) {\textlang{IncFst}};
\node[inner,  fill=innerB,  draw=borderB] (B1) at (\Bx, -0.625)
  {\textlang{example2()} \\ $[y \stackptr 1]$};
 
\draw[llarrow] (Bh.west) -- (Ah.east);
\node[draw, rounded corners=2pt, fill=gray!15, font=\scriptsize\ttfamily, minimum height=0.4cm]
  (head) at (\Bx + 2.25, 0) {head};
\draw[llarrow] (head.west) -- (Bh.east);
  
\end{tikzpicture}}
\end{center}
Above, we first notice how, contrary the fiber configurations in \Cref{sec:background-fibers}, 
we have regions inside the stack frames like we saw in \Cref{sec:background-regions}.
This fiber configuration is made up of two fibers: the head of the list points to the fiber installed by the 
\textlang{IncFst} effect, the \textlang{IncFst} fiber again points to the initial fiber 
that always marks the end of the stack. Observe how the two entries in the pair \textlang{p} 
used as argument to the effect \textlang{IncFst}
are references allocated in different regions: \textlang{x} is allocated 
in the region created by \textlang{handle\_inc\_fst}, and the reference \textlang{y} is allocated in 
the region created by the \textlang{example2} function. Further, the \textlang{example2} function is located in the fiber installed 
by the \textlang{IncFst} effect handler; this follows the structure of the program in \Cref{fig:inc-fst-impl} where 
\textlang{example2} is called after the \textlang{IncFst} effect handler is installed.
When \langkw{do} \textlang{IncFst p} is executed on line 6, and control is transferred to the effect handler, 
the fiber configuration changes to this: 
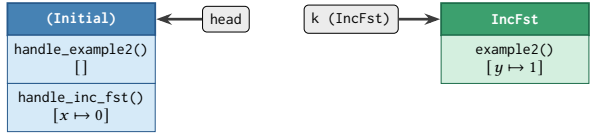
\begin{center}
  \scalebox{0.85}{\begin{tikzpicture}[
  header/.style = {
    draw, thick,
    minimum width  = 2.3cm,
    minimum height = 0.5cm,
    inner sep      = 4pt,
    text centered,
    font = \scriptsize\bfseries,
    text = white
  },
  inner/.style = {
    draw,
    minimum width  = 2.3cm,
    minimum height = 0.75cm,
    inner sep      = 3pt,
    text centered,
    align = center,
    font  = \scriptsize
  },
  llarrow/.style = {
    -{Stealth[length=7pt, width=5pt]},
    line width = 1.2pt,
    color = arrowC
  }
]
 
\def\Ax{0}
\node[header, fill=headerA, draw=borderA] (Ah) at (\Ax,  0.000) {\textlang{(Initial)}};
\node[inner,  fill=innerA,  draw=borderA] (A1) at (\Ax, -0.625)
  {\textlang{handle\_example2()} \\ $[]$};
\node[inner,  fill=innerA,  draw=borderA] (A2) at (\Ax, -1.375)
  {\textlang{handle\_inc\_fst()} \\ $[x \stackptr 0]$};
 
\def\Bx{6.75}
\node[header, fill=headerB, draw=borderB] (Bh) at (\Bx,  0.000) {\textlang{IncFst}};
\node[inner,  fill=innerB,  draw=borderB] (B1) at (\Bx, -0.625)
  {\textlang{example2()} \\ $[y \stackptr 1]$};
 
\node[draw, rounded corners=2pt, fill=gray!15, font=\scriptsize\ttfamily, minimum height=0.4cm]
  (head) at (\Ax + 2.25, 0) {head};
\draw[llarrow] (head.west) -- (Ah.east);
\node[draw, rounded corners=2pt, fill=gray!15, font=\scriptsize\ttfamily, minimum height=0.4cm]
  (cont) at (\Bx - 2.55, 0) {\textlang{k (IncFst)}};
\draw[llarrow] (cont.east) -- (Bh.west);
  
\end{tikzpicture}}
\end{center}
The head now points to the initial fiber, and the continuation available inside the handler branch 
of the \textlang{IncFst} effect points to the \textlang{IncFst} fiber.
As the handler branch increments the reference \textlang{x}, through the first entry in the  pair it receives, 
it executes safely; 
the reference \textlang{x} sits in a region which is a part of the current stack. 
We deem it undefined behavior to use local references that are not part of the current stack, 
\ie local references allocated in a region of a fiber that a continuation points to
and not a fiber from the list of fibers that make up the current stack,
such as if the increment had instead been done on the reference \textlang{y} through the second entry in the pair.
We classify this as undefined behavior for two reasons: 
(1) to ensure that references can only be used inside the scope of the region they belong to; and
(2) to ensure that reclamation of memory used by fibers of continuations should not depend
on anything except on how the continuation itself is used.

After the handler branch of \textlang{IncFst} is executed, 
and the computation is resumed by using the continuation, the fiber configuration is this: 
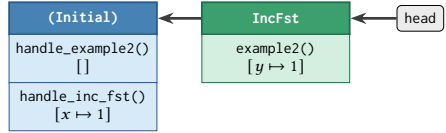
\begin{center}
  \scalebox{0.85}{\begin{tikzpicture}[
  header/.style = {
    draw, thick,
    minimum width  = 2.3cm,
    minimum height = 0.5cm,
    inner sep      = 4pt,
    text centered,
    font = \scriptsize\bfseries,
    text = white
  },
  inner/.style = {
    draw,
    minimum width  = 2.3cm,
    minimum height = 0.75cm,
    inner sep      = 3pt,
    text centered,
    align = center,
    font  = \scriptsize
  },
  llarrow/.style = {
    -{Stealth[length=7pt, width=5pt]},
    line width = 1.2pt,
    color = arrowC
  }
]
 
\def\Ax{0}
\node[header, fill=headerA, draw=borderA] (Ah) at (\Ax,  0.000) {\textlang{(Initial)}};
\node[inner,  fill=innerA,  draw=borderA] (A1) at (\Ax, -0.625)
  {\textlang{handle\_example2()} \\ $[]$};
\node[inner,  fill=innerA,  draw=borderA] (A2) at (\Ax, -1.375)
  {\textlang{handle\_inc\_fst()} \\ $[x \stackptr 1]$};
 
\def\Bx{3.0}
\node[header, fill=headerB, draw=borderB] (Bh) at (\Bx,  0.000) {\textlang{IncFst}};
\node[inner,  fill=innerB,  draw=borderB] (B1) at (\Bx, -0.625)
  {\textlang{example2()} \\ $[y \stackptr 1]$};
 
\draw[llarrow] (Bh.west) -- (Ah.east);
\node[draw, rounded corners=2pt, fill=gray!15, font=\scriptsize\ttfamily, minimum height=0.4cm]
  (head) at (\Bx + 2.25, 0) {head};
\draw[llarrow] (head.west) -- (Bh.east);
  
\end{tikzpicture}}
\end{center}
We are back to the configuration we had before \langkw{do} \textlang{IncFst p} was executed on line 6, 
except the handler branch of \textlang{IncFst} has been executed in the meantime which results in \textlang{x} 
now storing \textlang{1} inside the region of the \textlang{handle\_inc\_fst} function. 
This fiber configuration implies that the assertion on line 6 holds.

\paragraph{Multi-shot effects} For multi-shot effects, we can not resume 
computation using continuations that capture regions as part of the fibers they point to
and keep using the reference allocated in these regions. To illustrate this, 
we reconsider the example with the multi-shot effect \textlang{Choose} 
and the one-shot effect \textlang{State} from \Cref{fig:choose-state}, 
but this time we also take into account the code in the comments for regions and modes.
The reference \textlang{r} inside the handler for the \textlang{State} effect on line 6 is now a local reference. 
We use the \textlang{\onceMode} mode for the \textlang{State} effect and the \textlang{\manyMode} mode for 
the \textlang{Choose} effect, as they are respectively one-shot and multi-shot effects. 
Just before the \textlang{Choose} effect is performed on line 12, the fiber configuration is shown below:
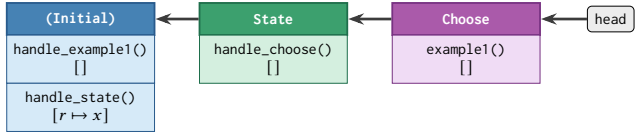
\begin{center}
  \scalebox{0.85}{\begin{tikzpicture}[
  header/.style = {
    draw, thick,
    minimum width  = 2.3cm,
    minimum height = 0.5cm,
    inner sep      = 4pt,
    text centered,
    font = \scriptsize\bfseries,
    text = white
  },
  inner/.style = {
    draw,
    minimum width  = 2.3cm,
    minimum height = 0.75cm,
    inner sep      = 3pt,
    text centered,
    align = center,
    font  = \scriptsize
  },
  llarrow/.style = {
    -{Stealth[length=7pt, width=5pt]},
    line width = 1.2pt,
    color = arrowC
  }
]
 
\def\Ax{0}
\node[header, fill=headerA, draw=borderA] (Ah) at (\Ax,  0.000) {\textlang{(Initial)}};
\node[inner,  fill=innerA,  draw=borderA] (A1) at (\Ax, -0.625)
  {\textlang{handle\_example1()} \\ $[]$};
\node[inner,  fill=innerA,  draw=borderA] (A2) at (\Ax, -1.375)
  {\textlang{handle\_state()} \\ $[r \stackptr x]$};
 
\def\Bx{3.0}
\node[header, fill=headerB, draw=borderB] (Bh) at (\Bx,  0.000) {\textlang{State}};
\node[inner,  fill=innerB,  draw=borderB] (B1) at (\Bx, -0.625)
  {\textlang{handle\_choose()} \\ $[]$};

\def\Cx{6.0}
\node[header, fill=headerC, draw=borderC] (Ch) at (\Cx,  0.000) {\textlang{Choose}};
\node[inner,  fill=innerC,  draw=borderC] (C1) at (\Cx, -0.625)
  {\textlang{example1()} \\ $[]$};
 
\draw[llarrow] (Bh.west) -- (Ah.east);
\draw[llarrow] (Ch.west) -- (Bh.east);
\node[draw, rounded corners=2pt, fill=gray!15, font=\scriptsize\ttfamily, minimum height=0.4cm]
  (head) at (\Cx + 2.25, 0) {head};
\draw[llarrow] (head.west) -- (Ch.east);
  
\end{tikzpicture}}
\end{center}
This is similar to the fiber configuration we saw in \Cref{sec:background-fibers} when 
we also considered the closed example \textlang{handle\_example1}. 
This time regions are included in the fiber configuration which illustrates
that the reference used to implement the \textlang{State} effect sits inside 
the region of the \textlang{handle\_state} function. When the \textlang{Choose} effect 
is performed on line 12, and control is suspended to the effect handler in \textlang{handle\_choose}, 
the fiber configuration changes to this: 
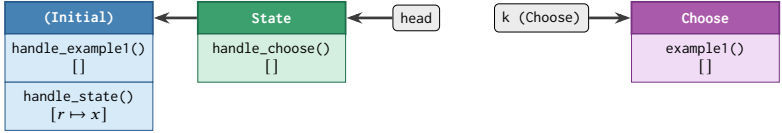
\begin{center}
  \scalebox{0.85}{\begin{tikzpicture}[
  header/.style = {
    draw, thick,
    minimum width  = 2.3cm,
    minimum height = 0.5cm,
    inner sep      = 4pt,
    text centered,
    font = \scriptsize\bfseries,
    text = white
  },
  inner/.style = {
    draw,
    minimum width  = 2.3cm,
    minimum height = 0.75cm,
    inner sep      = 3pt,
    text centered,
    align = center,
    font  = \scriptsize
  },
  llarrow/.style = {
    -{Stealth[length=7pt, width=5pt]},
    line width = 1.2pt,
    color = arrowC
  }
]
 
\def\Ax{0}
\node[header, fill=headerA, draw=borderA] (Ah) at (\Ax,  0.000) {\textlang{(Initial)}};
\node[inner,  fill=innerA,  draw=borderA] (A1) at (\Ax, -0.625)
  {\textlang{handle\_example1()} \\ $[]$};
\node[inner,  fill=innerA,  draw=borderA] (A2) at (\Ax, -1.375)
  {\textlang{handle\_state()} \\ $[r \stackptr x]$};
 
\def\Bx{3.0}
\node[header, fill=headerB, draw=borderB] (Bh) at (\Bx,  0.000) {\textlang{State}};
\node[inner,  fill=innerB,  draw=borderB] (B1) at (\Bx, -0.625)
  {\textlang{handle\_choose()} \\ $[]$};

\def\Cx{9.75}
\node[header, fill=headerC, draw=borderC] (Ch) at (\Cx,  0.000) {\textlang{Choose}};
\node[inner,  fill=innerC,  draw=borderC] (C1) at (\Cx, -0.625)
  {\textlang{example1()} \\ $[]$};
 
\draw[llarrow] (Bh.west) -- (Ah.east);
\node[draw, rounded corners=2pt, fill=gray!15, font=\scriptsize\ttfamily, minimum height=0.4cm]
  (head) at (\Bx + 2.25, 0) {head};
\draw[llarrow] (head.west) -- (Bh.east);
\node[draw, rounded corners=2pt, fill=gray!15, font=\scriptsize\ttfamily, minimum height=0.4cm]
  (cont) at (\Cx - 2.55, 0) {\textlang{k (Choose)}};
\draw[llarrow] (cont.east) -- (Ch.west);
  
\end{tikzpicture}}
\end{center}
The continuation of \textlang{Choose} captures a fiber with an empty region, but this is not always the case;
suppose the \textlang{Choose} effect handler is instead installed before 
the \textlang{State} handler as in the code below: 
\begin{lstlisting}
let handle_unsafe () = region (handle_choose (fun () => handle_state 0 example1))
\end{lstlisting}
Before performing the \textlang{Choose} effect in \textlang{example1} using \textlang{handle\_unsafe}, 
the fiber configuration looks like this: 
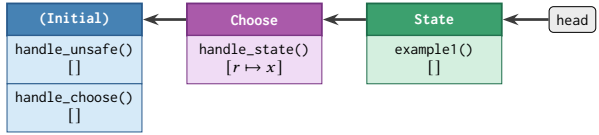
\begin{center}
  \scalebox{0.85}{\begin{tikzpicture}[
  header/.style = {
    draw, thick,
    minimum width  = 2.1cm,
    minimum height = 0.5cm,
    inner sep      = 4pt,
    text centered,
    font = \scriptsize\bfseries,
    text = white
  },
  inner/.style = {
    draw,
    minimum width  = 2.1cm,
    minimum height = 0.75cm,
    inner sep      = 3pt,
    text centered,
    align = center,
    font  = \scriptsize
  },
  llarrow/.style = {
    -{Stealth[length=7pt, width=5pt]},
    line width = 1.2pt,
    color = arrowC
  }
]
 
\def\Ax{0}
\node[header, fill=headerA, draw=borderA] (Ah) at (\Ax,  0.000) {\textlang{(Initial)}};
\node[inner,  fill=innerA,  draw=borderA] (A1) at (\Ax, -0.625)
  {\textlang{handle\_unsafe()} \\ $[]$};
\node[inner,  fill=innerA,  draw=borderA] (A2) at (\Ax, -1.375)
  {\textlang{handle\_choose()} \\ $[]$};
 
\def\Bx{2.8}
\node[header, fill=headerC, draw=borderC] (Bh) at (\Bx,  0.000) {\textlang{Choose}};
\node[inner,  fill=innerC,  draw=borderC] (B1) at (\Bx, -0.625)
  {\textlang{handle\_state()} \\ $[r \stackptr x]$};

\def\Cx{5.6}
\node[header, fill=headerB, draw=borderB] (Ch) at (\Cx,  0.000) {\textlang{State}};
\node[inner,  fill=innerB,  draw=borderB] (C1) at (\Cx, -0.625)
  {\textlang{example1()} \\ $[]$};

\draw[llarrow] (Bh.west) -- (Ah.east);
\draw[llarrow] (Ch.west) -- (Bh.east);
\node[draw, rounded corners=2pt, fill=gray!15, font=\scriptsize\ttfamily, minimum height=0.4cm]
  (head) at (\Cx + 2.15, 0) {head};
\draw[llarrow] (head.west) -- (Ch.east);
\end{tikzpicture}}
\end{center}
This fiber configuration reflects that we have changed the order in which the effect handlers are installed.
When the \textlang{Choose} effect is performed in \textlang{example1}, 
the fiber configuration changes to the one below 
where the continuation \textlang{k} in the \textlang{Choose} effect handler 
now captures both the \textlang{Choose} and \textlang{State} fiber which includes a non-empty region:
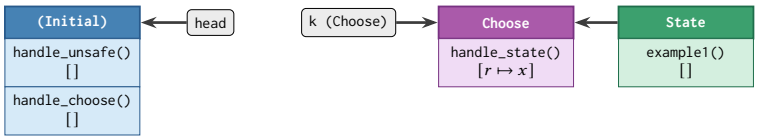
\begin{center}
  \scalebox{0.85}{\begin{tikzpicture}[
  header/.style = {
    draw, thick,
    minimum width  = 2.1cm,
    minimum height = 0.5cm,
    inner sep      = 4pt,
    text centered,
    font = \scriptsize\bfseries,
    text = white
  },
  inner/.style = {
    draw,
    minimum width  = 2.1cm,
    minimum height = 0.75cm,
    inner sep      = 3pt,
    text centered,
    align = center,
    font  = \scriptsize
  },
  llarrow/.style = {
    -{Stealth[length=7pt, width=5pt]},
    line width = 1.2pt,
    color = arrowC
  }
]
 
\def\Ax{0}
\node[header, fill=headerA, draw=borderA] (Ah) at (\Ax,  0.000) {\textlang{(Initial)}};
\node[inner,  fill=innerA,  draw=borderA] (A1) at (\Ax, -0.625)
  {\textlang{handle\_unsafe()} \\ $[]$};
\node[inner,  fill=innerA,  draw=borderA] (A2) at (\Ax, -1.375)
  {\textlang{handle\_choose()} \\ $[]$};

\def\Bx{6.75}
\node[header, fill=headerC, draw=borderC] (Bh) at (\Bx,  0.000) {\textlang{Choose}};
\node[inner,  fill=innerC,  draw=borderC] (B1) at (\Bx, -0.625)
  {\textlang{handle\_state()} \\ $[r \stackptr x]$};

\def\Cx{9.55}
\node[header, fill=headerB, draw=borderB] (Ch) at (\Cx,  0.000) {\textlang{State}};
\node[inner,  fill=innerB,  draw=borderB] (C1) at (\Cx, -0.625)
  {\textlang{example1()} \\ $[]$};

\draw[llarrow] (Ch.west) -- (Bh.east);
\node[draw, rounded corners=2pt, fill=gray!15, font=\scriptsize\ttfamily, minimum height=0.4cm]
  (head) at (\Ax + 2.15, 0) {head};
\draw[llarrow] (head.west) -- (Ah.east);
\node[draw, rounded corners=2pt, fill=gray!15, font=\scriptsize\ttfamily, minimum height=0.4cm]
  (cont) at (\Bx - 2.45, 0) {\textlang{k (Choose)}};
\draw[llarrow] (cont.east) -- (Bh.west);
\end{tikzpicture}}
\end{center}
The local reference inside the region of \textlang{handle\_state} is implicitly deallocated when the multi-shot 
effect \textlang{Choose} is performed. This is because the continuation \textlang{k} 
captures the region of \textlang{handle\_state}, and \textlang{k} is a multi-shot continuation; 
multi-shot continuations are created by making copies of the fibers they point to, but we can not make copies of regions 
and continue using the same local references inside these regions when computation is resumed 
as the underlying memory of the regions may have changed. 
When computation is resumed 
using the continuation \textlang{k} in the effect handler for \textlang{Choose}, we return 
to the fiber configuration we had before \textlang{Choose} was performed inside \textlang{example1} on line 
12 in \Cref{fig:choose-state}, except the reference \textlang{r} is no longer useable:
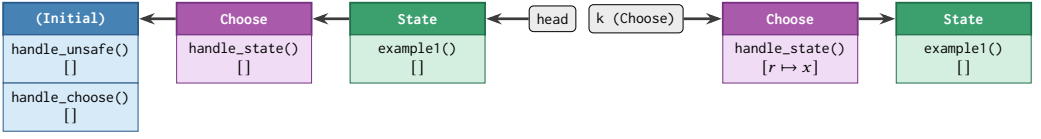
\begin{center}
  \scalebox{0.85}{\begin{tikzpicture}[
  header/.style = {
    draw, thick,
    minimum width  = 2.1cm,
    minimum height = 0.5cm,
    inner sep      = 4pt,
    text centered,
    font = \scriptsize\bfseries,
    text = white
  },
  inner/.style = {
    draw,
    minimum width  = 2.1cm,
    minimum height = 0.75cm,
    inner sep      = 3pt,
    text centered,
    align = center,
    font  = \scriptsize
  },
  llarrow/.style = {
    -{Stealth[length=7pt, width=5pt]},
    line width = 1.2pt,
    color = arrowC
  }
]
 
\def\Ax{0}
\node[header, fill=headerA, draw=borderA] (Ah) at (\Ax,  0.000) {\textlang{(Initial)}};
\node[inner,  fill=innerA,  draw=borderA] (A1) at (\Ax, -0.625)
  {\textlang{handle\_unsafe()} \\ $[]$};
\node[inner,  fill=innerA,  draw=borderA] (A2) at (\Ax, -1.375)
  {\textlang{handle\_choose()} \\ $[]$};
 
\def\Bx{2.7}
\node[header, fill=headerC, draw=borderC] (Bh) at (\Bx,  0.000) {\textlang{Choose}};
\node[inner,  fill=innerC,  draw=borderC] (B1) at (\Bx, -0.625)
  {\textlang{handle\_state()} \\ $[]$};

\def\Cx{5.4}
\node[header, fill=headerB, draw=borderB] (Ch) at (\Cx,  0.000) {\textlang{State}};
\node[inner,  fill=innerB,  draw=borderB] (C1) at (\Cx, -0.625)
  {\textlang{example1()} \\ $[]$};

\def\Dx{11.2}
\node[header, fill=headerC, draw=borderC] (Dh) at (\Dx,  0.000) {\textlang{Choose}};
\node[inner,  fill=innerC,  draw=borderC] (D1) at (\Dx, -0.625)
  {\textlang{handle\_state()} \\ $[r \stackptr x]$};

\def\Ex{13.9}
\node[header, fill=headerB, draw=borderB] (Eh) at (\Ex,  0.000) {\textlang{State}};
\node[inner,  fill=innerB,  draw=borderB] (E1) at (\Ex, -0.625)
  {\textlang{example1()} \\ $[]$};

\draw[llarrow] (Bh.west) -- (Ah.east);
\draw[llarrow] (Ch.west) -- (Bh.east);
\node[draw, rounded corners=2pt, fill=gray!15, font=\scriptsize\ttfamily, minimum height=0.4cm]
  (head) at (\Cx + 2.10, 0) {head};
\draw[llarrow] (head.west) -- (Ch.east);
\draw[llarrow] (Dh.east) -- (Eh.west);
\node[draw, rounded corners=2pt, fill=gray!15, font=\scriptsize\ttfamily, minimum height=0.4cm]
  (cont) at (\Dx - 2.40, 0) {\textlang{k (Choose)}};
\draw[llarrow] (cont.east) -- (Dh.west);
\end{tikzpicture}}
\end{center}
This fiber configuration implies that it is unsafe to use the \textlang{State} effect on lines 19 and 20 
in the \textlang{example1} function when using \textlang{handle\_unsafe}, as the reference \textlang{r} is no longer 
useable as part of the list of fibers that makes up the current stack 
(the data \textlang{x} may still be in the region of \textlang{handle\_state} at ruuntime but not at the location \textlang{r} represents)
\paragraph{Skippable detour about references and multi-shot continuations}
References captured by multi-shot continuations can be reused under certain limiting restrictions. 
For example, \citet{Muhcu25} describes a runtime for a language with multi-shot effects where continuations can capture 
stack reference as part of their closures by using an indirection layer for references.
In that setting, a stack reference points to a reference in an indirection layer which again points to the actual 
stack. This approach \citep{Muhcu25}, however, only works as long as no two continuations ever capture the same fiber.
If we instead took the approach of reusing the same region, or parts of it, for multiple 
fibers to preserve the local references,  
we could no longer implement true region-based memory management where the memory underlying regions 
is freed when code exits the scope of a region; a region can be exited more than once,
making it unsafe for all invocations of a continuation, apart from the very first, to use memory in regions.
\subsection{Syntax and Operational Semantics}
\label{sec:semantics}
In this subsection, we formalize the operational semantics for the runtime model we gave an overview of in the previous 
subsection. We start by defining the expressions and values of our language below, 
where we omit products, sums, unary operators and binary operators for brevity.
\begin{align*}
  \vInt \in & ~ \mathds{Z} \qquad
  \var \in ~ \Var \qquad
  \loc \in \Loc \qquad
  \functionid \in \Functionid \qquad
  \ope \in \Operation \\
  \addrVar \in \Addresses \bnfdef{} &
   \saddr{\loc} 
   \ALT \haddr{\loc} \qquad
  \localVar \in \LocalMode \bnfdef{} 
   \localMode 
    \ALT \globalMode \qquad
  \affineVar \in \AffineMode \bnfdef{}
    \onceMode 
    \ALT \manyMode \\
  \val \in \Val \bnfdef{}
  & \TT
    \ALT \vInt
    \ALT \var
    \ALT \True
    \ALT \False
    \ALT \addrVar
    \ALT \vlambda{\addrVar}{\functionid}{f}{\var}{\expr} 
    \ALT \contOne{\addrVar}{\functionid}{\lctx}{\fibers}{\loc}
    \ALT \contMul{\addrVar}{\functionid}{\lctx}{\fibers} 
    \ALT \dots \\
  \expr \in \Expr \bnfdef{} 
  & \val
    \ALT \var
    \ALT \Let \var = \expr_{1} in \expr_{2}
    \ALT \expr_1 ~ \expr_2
    \ALT \If \expr_{1} then \expr_{2} \Else \expr_{3}
    \ALT \ealloc{\localVar}{\expr}
    \ALT \eload{\expr}
    \ALT \estore{\expr_1}{\expr_2} \\
    & \ALT (\ematch{\expr_1}{\var_1}{\expr_1}{\var_2}{\expr_2})
    \ALT \elambda{\localVar}{f}{\var}{\expr} 
    \ALT \eregion{\expr}
    \ALT \eend{\expr} \\
    & \ALT \edo{\ope}{\expr} 
    \ALT \eeffect{\ope}{\val}{\lctx} 
    \ALT (\ehandle{\localVar}{\affineVar}{\expr_1}{\ope}{\var}{\con}{\expr_2}{\var}{\expr_3}) \\
    & \ALT (\einstall{\localVar}{\affineVar}{\expr_1}{\ope}{\var}{\con}{\expr_2}{\var}{\expr_3})
    \ALT \dots
\end{align*}
\paragraph{Expressions and values}
The values in the language include the unit value, integers, booleans, addresses, $\lambda$-abstractions, 
one-shot continuations and multi-shot continuations.
Addresses are either stack addresses $\saddr{\loc}$, with a logical location $\loc$, 
or heap address $\haddr{\loc}$ likewise with a location.
In $\lambda$-abstractions, $f$ is the recursive occurrence in the body $\expr$, $x$ is an argument and 
the superscript pair $(\addrVar, \functionid)$ says that the closure of the abstraction, represented with 
closure identifier $\functionid$, is located at address $\addrVar$. This way of tracking 
closures is similar to \citet{Georges25}, even though \citeauthor{Georges25} formalize a language without effect handlers.
For our continuation values, we use the same mechanism for tracking closures. 
The value for one-shot continuations $\contOne{\addrVar}{\functionid}{\lctx}{\fibers}{\loc}$, 
takes an evaluation context $\lctx$ (which we define momentarily), a list of fibers $\fibers$ 
(which we also define momentarily) and a location $\loc$ that is used to enforce the affinity of one-shot continuations.
The value for multi-shot continuation $\contMul{\addrVar}{\functionid}{\lctx}{\fibers}$ 
is like the value for one-shot continuations, except it does not have the location that we use to enforce affinity. 

The expressions in our language include values, variables, conditionals, sequencing, function application, 
allocation of local references in regions and global references on the heap, 
and expressions for storing and loading from references. 
In addition, we have the expression version of $\lambda$-abstractions $\elambda{\localVar}{f}{\var}{\expr}$; 
this is used to allocate function closures, 
the local mode $\localVar$ in superscript determines whether the closure is allocated in a region or on the heap.
Further, we have the $\eregion{\expr}$ expression for starting a region around an expression $\expr$,
and $\eend{\expr}$ which we use to mark when the scope of a region ends and its underlying memory is freed.
Lastly, we have a number of expressions for algebraic effects: we saw the expression 
$\ehandle{\localVar}{\affineVar}{\expr_1}{\ope}{\var}{\con}{\expr_2}{\var}{\expr_3}$ throughout \Cref{sec:overview} 
for installing an effect handler and its underlying fiber. 
Contrary to the previous work on program logics for effect handlers in Iris \citep{de2021separation},
our \langkw{try}-expression is parameterized with a pair of modes for controlling 
whether the effect is one-shot or multi-shot, and if the closure of the continuation $k$ is allocated locally 
in a region or on the heap. Similarly to how  $\eend{\expr}$ marks the end of a region, 
$\einstall{\localVar}{\affineVar}{\expr_1}{\ope}{\var}{\con}{\expr_2}{\var}{\expr_3}$ marks the end of a fiber:
when the expression $\expr_1$ evaluates to a value, the return branch $\expr_3$ is executed with the value as argument, 
and at the same time, the fiber with operation $\ope$ that the \langkw{try}-expression installed is removed from the list of fibers 
that make up the stack. When an effect is performed using $\edo{\ope}{\expr}$, 
we use the mechanism from \citet{de2021separation} to capture the paused 
computation in continuation values, this is what the $\eeffect{\ope}{\val}{\lctx}$ expression is for. 
We will see exactly how this works when we present the reduction rules of our language 
which are based on \emph{evaluation contexts}. 
\paragraph{Evaluation contexts} 
We use evaluation contexts $\lctx$ below to define the language. 
$\bullet$ is the empty context, and $\lctx[\expr]$ is notation for filling out the evaluation context $\lctx$ with the expression $\expr$.
We list the evaluation contexts of our language below (again, we omit contexts for products, sums, unary operators and binary operators for brevity). 
\begin{align*}
 \nlctx' \in \NLctx' \bnfdef{}
  & \Let \var = \bullet in \expr
  \ALT \expr ~ \bullet
  \ALT \bullet ~ \val
  \ALT \If \bullet then \expr_{1} \Else \expr_{2} 
  \ALT \ealloc{\localVar}{\bullet}
  \ALT \eload{\bullet} \\ 
  & \ALT \estore{\expr}{\bullet} 
  \ALT \estore{\bullet}{\val} 
  \ALT (\ematch{\bullet}{\var_1}{\expr_1}{\var_2}{\expr_2}) \\
  & \ALT \eend{\bullet} 
  \ALT \edo{\ope}{\bullet} 
  \ALT (\einstall{\localVar}{\affineVar}{\bullet}{\ope}{\var}{\con}{\expr_2}{\var}{\expr_3})
  \ALT \dots \\
  \lctx' \in \Lctx' \bnfdef{} 
  & \nlctx'
  \ALT (\ehandle{\localVar}{\affineVar}{\bullet}{\ope}{\var}{\con}{\expr_1}{\var}{\expr_2}) \\
  \nlctx \in \NLctx \bnfdef{} 
  & \bullet
  \ALT \nlctx'[\nlctx]
  \qquad 
  \lctx \in \Lctx \bnfdef{} 
  \bullet
  \ALT \lctx'[\lctx]
\end{align*}
We say that all evaluation contexts except those for the \langkw{try}-expression are \emph{neutral contexts}. 
We need neutrals contexts for when we capture delimited continuations using the $\eeffect{\ope}{\val}{\lctx}$ expression in the reduction rules. 
\paragraph{Reduction rules} 
The program state ($\heap$, $\fibers$, $\locations$, $\functions$) used in our reduction 
rules is defined as follows:
\begin{align*}
  \heap \in \Heap & \eqdef{} \Loc \fpfn (\Val + ~ \Functionid) \qquad
  \locations \in \Locations\eqdef{} \mathcal{P}_{\text{fin}}(\Loc) \qquad 
  \functions \in \Functions \eqdef{} \mathcal{P}_{\text{fin}}(\Functionid) \\
  \sframe \in \StackFrame & \eqdef{} \text{List} ~ (\Loc \times (\Val + ~ \Functionid)) \qquad
  \sframes \in \StackFrames \eqdef{} \text{List} ~ \StackFrame \\
  \fibers \in \Fibers & \eqdef{} \text{List} ~ (\Operation \times ~ \StackFrames) \qquad
  \progstate \in \Progstate{} \eqdef{} \Heap \times \Fibers \times \Locations \times \Functions
\end{align*}
The state consists of the heap $\heap$, the installed fibers $\fibers$ that make up the stack, a set of locations $\locations$ and a set of closure identifiers $\functions$.
In the definition of the heap $\heap$ above, we see how the heap consists of both values and closure identifiers. 
We keep around a set $\functions$ in the program state that tracks all the previous used closure identifiers 
to make sure we are always generating new identifiers when allocating closures. 
Fibers, like the heap, can contain both values and closure identifiers. Fibers are built using a number of layers: 
we retain the association of regions with stack frames that we used throughout \Cref{sec:overview}. 
Thus, fibers are defined as a list of stack frames, a stack frame is a list of locations paired with either the value or the closure 
identifier that it stores. The set of logical locations $\locations$ tracks all the locations that have previously been used for local 
allocations. When making a local allocation, we can not rely on the fibers in the program state alone when we want to generate a fresh location;
continuation values also hold fibers with locations in them. Thus, we use the set $\locations$ for new local locations.
Having defined the program state, we can define the reduction relations for the operational semantics of Yarrow.

Our small-step operational semantics consists of two reduction relations, the base step ($\bstep$)
and the context step ($\cstep$), both with signature $\Expr \times \Progstate \cstep \Expr \times \Progstate$.
The context step only has one reduction rule shown below: 
\begin{align*}
\inferH{Evaluation-Context-Step}
  {(\expr, \progstate) \bstep (\expr', \progstate')}
  {(\lctx[\expr], \progstate) \cstep (\lctx[\expr'], \progstate')}
\end{align*}
This rule evaluates an expression in an evaluation context $\lctx[\expr]$ to $\lctx[\expr']$, 
if there is a base step reduction from $\expr$ to $\expr'$. 
In \Cref{fig:non-eff-base-steps}, we display an excerpt of base step reduction rules unrelated to 
algebraic effects, 
and in \Cref{fig:eff-base-steps} we display another excerpt that are about algebraic effects.
First, we focus on the reduction rules that are not about algebraic effects.
\begin{figure}[]
  \centering
  \begin{mathpar}
  \axiomH{Region}
  {\big(\eregion{\expr}, (\heap, \fibers \mdoubleplus [(\ope, ~ \sframes)], \locations, \functions)\big) \bstep 
   \big(\eend{\expr}, (\heap, \fibers \mdoubleplus [(\ope, ~ \sframes \mdoubleplus [[]])], \locations, \functions)\big)\big)}
  \and
  \axiomH{End}
  {\big(\eend{\val},(\heap, \fibers \mdoubleplus [(\ope, ~ \sframes \mdoubleplus [\sframe])], \locations, \functions)\big) \bstep 
    \big(\val, (\heap, \fibers \mdoubleplus [(\ope, ~ \sframes)], \locations, \functions)\big)}
  \and
  \inferH{Alloc-Global}
  {\loc ~ \text{fresh in} ~ \heap}
  {\big(\ealloc{\globalMode}{\val}, (\heap, \fibers, \locations, \functions)\big) \bstep 
   \big(\haddr{\loc}, (\heap[\loc := \val], \fibers, \locations, \functions)\big)}
  \and
  \inferH{Alloc-Local}
  {\loc ~ \text{fresh in} ~ \locations}
  {\big(\ealloc{\localMode}{\val}, (\heap, \fibers \mdoubleplus [(\ope, ~ \sframes \mdoubleplus [\sframe])], \locations, \functions)\big) \bstep
  \big(\saddr{\loc}, (\heap, \fibers \mdoubleplus [(\ope, ~ \sframes \mdoubleplus [\sframe \mdoubleplus [(\loc, \val)]])], \locations \uplus \{\loc\},\functions)\big)}
  \and
  \axiomH{Load-Global}
  {\big(\eload{(\haddr{\loc})}, (\heap[\loc := \val], \fibers, \locations, \functions)\big) \bstep 
   \big(\val, (\heap[\loc := \val], \fibers, \locations, \functions)\big)}
  \and
  \axiomH{Store-Local}
  {\big(\estore{(\saddr{\loc})}{\val_1}, (\heap, \fibers_1 \mdoubleplus [(\ope, ~ \sframes_1 \mdoubleplus [\sframe_1 \mdoubleplus [(\loc, \val_2)] \mdoubleplus \sframe_2] \mdoubleplus \sframes_2)] \mdoubleplus \fibers_2, \locations, \functions)\big) \bstep \\
  \big(\TT, (\heap, \fibers_1 \mdoubleplus [(\ope, ~ \sframes_1 \mdoubleplus [\sframe_1 \mdoubleplus [(\loc, \val_1)] \mdoubleplus \sframe_2] \mdoubleplus \sframes_2)] \mdoubleplus \fibers_2, \locations, \functions)\big)}
  \and
  \inferH{Function-Global}
  {\loc ~ \text{fresh in} ~ \heap \and \functionid ~ \text{fresh in} ~ \functions}
  {\big((\elambda{\globalMode}{f}{\var}{\expr}), (\heap, \fibers, \locations, \functions)\big) \bstep
   \big((\vlambda{\haddr{\loc}}{\functionid}{f}{\var}{\expr}), (\heap[\loc := \functionid], \fibers, \locations, \functions \uplus \{\functionid\})\big)}
  \and
  \inferH{Function-Local}
  {\loc ~ \text{fresh in} ~ \locations \and \functionid ~ \text{fresh in} ~ \functions}
  {\big((\elambda{\localMode}{f}{\var}{\expr}), (\fibers \mdoubleplus [(\ope, ~ \sframes \mdoubleplus [\sframe])], \locations, \functions)\big) \bstep \\
   \big((\vlambda{\saddr{\loc}}{\functionid}{f}{\var}{\expr}), (\heap, \fibers \mdoubleplus [(\ope, ~ \sframes \mdoubleplus [\sframe \mdoubleplus [(\loc, \functionid)]])], \locations \uplus \{\loc\}, \functions \uplus \{\functionid\})\big)}
  \and
  \inferH{Application}
  {(\addrVar, \functionid) ~ \text{alive in} ~ \progstate}
  {\big((\vlambda{\loc}{\functionid}{f}{\var}{\expr}) ~ \val, \progstate\big) \bstep \big(\expr[(\vlambda{\loc}{\functionid}{f}{\var}{\expr})/f][\val/\var], \progstate\big)}
  \end{mathpar}
  \caption{Selected base steps reduction rules.}
  \label{fig:non-eff-base-steps}
  \vspace{-1mm}
\end{figure}

In the \ruleref{Region} rule, the $\eregion{\expr}$ expression steps to the $\eend{\expr}$ expression, doing so, 
a new empty stack frame is inserted in the most recently installed fiber. 
When the expression $\expr$ in $\eend{\expr}$ evaluates to a value $\val$, 
the top most stack frame $\sframe$ is freed, in the \ruleref{End} rule, which 
effectively ends the region as the local references that reside in the stack frame $\sframe$ are no longer available.
Local references are always allocated into the top most stack frame which corresponds to 
the most recently started region (\ruleref{Alloc-Local}). Loads and stores to local references (\ruleref{Store-Local}) 
happen in place in the stack frame where the reference was allocated. 
Global heap references follow a standard semantics, \eg \ruleref{Alloc-Global} and \ruleref{Load-Global}. 
We also use references when allocating function closures:
In the \ruleref{Function-Global} and \ruleref{Function-Local} rules, the expression $\elambda{\localVar}{f}{\var}{\expr}$
steps to the value $\vlambda{\addrVar}{\functionid}{f}{\var}{\expr}$, 
$\functionid$ is a new identifier in the set of closure identifiers $\functions$ in the program state. 
The superscript $(\addrVar, \functionid)$ indicates that the address of the closure for this $\lambda$-abstaction
(the closure being represented by the closure id $\functionid$) is $\addrVar$.
The address $\addrVar$, which is either a heap address or a stack address 
depending on the locality mode $\localVar$, is set to point to the closure identifier.
When applying a $\lambda$-abstaction, we check in the antecedent of the \ruleref{Application} rule
whether the closure is still alive.
That is, the phrase "$(\addrVar, \functionid) ~ \text{alive in} ~ \progstate$"
means that the address $\addrVar$ points to $\functionid$ on the heap $\heap$ in $\progstate$, 
or the address $\addrVar$ points to $\functionid$ in any of the stack frames inside the fibers $\fibers$ in the state $\progstate$.
\begin{figure}[]
  \vspace{-1mm}
  \centering
  \begin{mathpar}
  \axiomH{Do}
  {\big(\edo{\ope}{\val}, \progstate\big) \bstep \big(\eeffect{\ope}{\val}{\bullet}, \progstate\big)}
  \and
  \axiomH{Eff}
  {\big(\nlctx[\eeffect{\ope}{\val}{\lctx}], \progstate\big) \bstep \big(\eeffect{\ope}{\val}{\nlctx[\lctx]}, \progstate\big)}
  \and
  \axiomH{Handler-Install}
  {\big(\ehandle{\localVar}{\affineVar}{\expr}{\ope}{\var}{\con}{\expr_1}{\var}{\expr_2}, (\heap, \fibers, \locations, \functions)\big) \bstep \\ 
   \big(\einstall{\localVar}{\affineVar}{\expr}{\ope}{\var}{\con}{\expr_1}{\var}{\expr_2}, (\heap, \fibers \mdoubleplus [(\ope, ~ [])], \locations, \functions)\big)}
  \and
  \inferH{Handler-NEQ}
  {\ope \neq \ope'}
  {\big(\einstall{\localVar}{\affineVar}{(\eeffect{\ope}{\val}{\lctx})}{\ope'}{\var}{\con}{\expr_1}{\var}{\expr_2}, \progstate\big) \bstep \\ 
   \big(\eeffect{\ope}{\val}{(\einstall{\localVar}{\affineVar}{\lctx}{\ope'}{\var}{\con}{\expr_1}{\var}{\expr_2})}, \progstate\big)}
  \and
  \inferH{Handler-Local-One}
  {\loc_1 ~ \text{fresh in} ~ \heap \and \loc_2 ~ \text{fresh in} ~ \locations \and \functionid ~ \text{fresh in} ~ \functions \and \ope ~ \text{not installed in} ~ \fibers_2}
  {\big(\einstall{\localMode}{\onceMode}{(\eeffect{\ope}{\val}{\lctx})}{\ope}{\var}{\con}{\expr_1}{\var}{\expr_2}, \\
    (\heap, \fibers_1 \mdoubleplus [(\ope', \sframes \mdoubleplus [\sframe])] \mdoubleplus [(\ope, \sframes)]\mdoubleplus \fibers_2, \locations, \functions)\big) \bstep \\ 
   \big(\expr_1[\val/\var][(\contOne{\saddr{\loc_2}}{\functionid}{(\einstall{\localMode}{\onceMode}{\lctx}{\ope}{\var}{\con}{\expr_1}{\var}{\expr_2})}{([(\ope, \sframes)] \mdoubleplus \fibers_2)}{\loc_1})/\con], \\ 
   (\heap[\loc_1 := \False], \fibers_1 \mdoubleplus [(\ope', \sframes \mdoubleplus [\sframe \mdoubleplus [(\loc_2, \functionid)]])], \locations \uplus \{\loc_2\}, \functions \uplus \{\functionid\})\big)}
  \and
  \inferH{Handler-Global-Multi}
  {\loc ~ \text{fresh in} ~ \heap \and \functionid ~ \text{fresh in} ~ \functions \and \ope ~ \text{not installed in} ~ \fibers_2}
  {\big(\einstall{\globalMode}{\manyMode}{(\eeffect{\ope}{\val}{\lctx})}{\ope}{\var}{\con}{\expr_1}{\var}{\expr_2}, 
    (\heap, \fibers_1 \mdoubleplus [(\ope, \sframes)]\mdoubleplus \fibers_2, \locations, \functions)\big) \bstep \\ 
   \big(\expr_1[\val/\var][(\contMul{\haddr{\loc}}{\functionid}{(\einstall{\globalMode}{\manyMode}{\lctx}{\ope}{\var}{\con}{\expr_1}{\var}{\expr_2})}{([(\ope, \sframes)] \mdoubleplus \fibers_2)})/\con], \\ 
   (\heap[\loc := \functionid], \fibers_1, \locations, \functions \uplus \{\functionid\})\big)}
  \and
  \inferH{Handler-Return}
  {}
  {\big(\einstall{\localVar}{\affineVar}{\val}{\ope}{\var}{\con}{\expr_1}{\var}{\expr_2}, (\heap, \fibers \mdoubleplus [(\ope, \sframes)], \locations, \functions)\big) \bstep
   \big(\expr_2[\val/\var], (\heap, \fibers, \locations, \functions)\big)}
  \and
  \inferH{Cont-One}
  {(\addrVar, \functionid) ~ \text{alive in} ~  (\heap, \fibers_1, \locations, \functions)}
  {\big((\contOne{\addrVar}{\functionid}{\lctx}{\fibers_2}{\loc}) ~ \val, (\heap[\loc := \False], \fibers_1, \locations, \functions)\big) \bstep 
   \big(\lctx[\val], (\heap[\loc := \True], \fibers_1 \mdoubleplus \fibers_2, \locations, \functions)\big)}
  \and
  \inferH{Cont-Multi}
  {(\loc, \functionid) ~ \text{alive in} ~ (\heap, \fibers_1, \locations, \functions)}
  {\big((\contMul{\addrVar}{\functionid}{\lctx}{\fibers_2}) ~ \val, (\heap, \fibers_1, \locations, \functions)) \bstep 
   \big(\lctx[\val], (\heap, \fibers_1 \mdoubleplus (rem\_locs ~ \fibers_2), \locations, \functions)\big)}
  \end{mathpar}
  \caption{Selected base step reduction rules for effects handlers.}
  \label{fig:eff-base-steps}
\end{figure}

Next we turn our attention to the reduction rules in \Cref{fig:eff-base-steps} about algebraic effects. 
The \ruleref{Handler-Install} rule installs an effect handler and its corresponding fiber by 
stepping from the \langkw{try}-expression to the corresponding \langkw{installed}-expression.
Initially, the newly installed fiber in the program state does not have any stack frames.
As mentioned earlier in this subsection, when an effect is performed using $\edo{\ope}{\val}$, the expression 
is replaced with $\eeffect{\ope}{\val}{\bullet}$ (see the \ruleref{Do} rule). From here, an evaluation context $\lctx$
is created in the $\eeffect{\ope}{\val}{\lctx}$ expression by consuming surrounding neutral context using the 
\ruleref{Eff} rule. If the evaluation context surrounding $\eeffect{\ope}{\val}{\lctx}$ is not a neutral context, 
it is by definition a \langkw{try}-context. When such a context is encountered, and the operation of the 
$\eeffect{\ope}{\val}{\lctx}$ expression is not the same as the operation in the \langkw{try}-context, we simply keep 
propagating upwards in the program, as shown in the \ruleref{Handler-NEQ} rule, until the correct effect handler is found. 
When the operation matches and the correct effect handler is found, we create a continuation value and 
substitute it, together with the argument provided when performing the effect, into the handler branch of the 
effect handler, see \ruleref{Handler-Global-Multi} and \ruleref{Handler-Local-One}.
For instance, in the \ruleref{Handler-Global-Multi} reduction rule, where the effect is multi-shot and the continuation closure is stored on the heap, 
the continuation value 
\begin{align*}
  \contMul{\haddr{\loc}}{\functionid}{(\einstall{\globalMode}{\manyMode}{\lctx}{\ope}{\var}{\con}{\expr_1}{\var}{\expr_2})}{([(\ope, \sframes)] \mdoubleplus \fibers_2)}
\end{align*}
is created. This value is made up of three parts: 
(1) the pair $(\haddr{\loc}, \functionid)$ stating that the continuation closure represented by the identifier $\functionid$ is stored in the heap location $\loc$
(2) an evaluation context consisting of $\lctx$, from the $\eeffect{\ope}{\val}{\lctx}$ expression, 
    around which the effect handler is reinstalled 
    (reinstalling the effect handler gives \emph{deep-handler} semantics contrary to \emph{shallow}-handlers where the effect handler is not reinstalled) 
(3) the fibers that the continuation captures, \ie{} all the fibers up to and including the fiber installed by effect handler. 
The \ruleref{Handler-Local-One} rule follows a similar pattern when creating a continuation value for a one-shot effect with a locally allocated 
closure, except the continuation closure is stored at a stack location $\loc_2$, and we allocate a heap location $\loc_1$ 
to enforce that the continuation is not called more than once:
\begin{align*}
\contOne{\saddr{\loc_2}}{\functionid}{(\einstall{\localMode}{\onceMode}{\lctx}{\ope}{\var}{\con}{\expr_1}{\var}{\expr_2})}{([(\ope, \sframes)] \mdoubleplus \fibers_2)}{\loc_1}.
\end{align*}
For brevity, we have omitted the reduction rules for multi-shot effects with local continuation closures and one-shot effects with global continuation closures, 
the rules we have shown (\ruleref{Handler-Global-Multi} and \ruleref{Handler-Local-One}) cover all the aspects of the rules.
When applying a one-shot continuation in the \ruleref{Cont-One} reduction rule, we make sure the continuation has not been 
used before by checking that the location $\loc$ points to false, whereafter we flip it to true, and that the continuation closure is still 
alive like in the \ruleref{Application} rule. afterwards, the continuation is called with a value $\val$ by placing $\val$ in the 
context of the continuation value $\lctx$ and pushing the fibers of the continuation value $\fibers_2$ onto the stack.
In \ruleref{Cont-Multi}, when a continuation of a multi-shot effect is called, we use the function $rem\_locs$ to remove the 
locations in the fibers $\fibers_2$, that is, $rem\_locs ~ \fibers_2$ contains all the same stack frames as $\fibers_2$, 
except the stack frames are empty. 
Because the stack frames are not removed, but emptied, the program $\lctx[v]$ that the continuation resumes, 
can allocate into and end the regions it had already started, but it can not use the references that were already allocated in these regions.
This models our approach to multi-shot continuations explained in \Cref{sec:overview}.


\section{Program Logic}
\label{sec:logic}
In this section, we present the Yarrow Logic (YL), an Iris-based program logic 
built to reason about the operational semantics of Yarrow we presented in \Cref{sec:semantics}.
YL inherits all the reasoning principles from Iris; 
we assume familiarity with separation logic and use this section to present the features 
that are unique to YL.
\subsection{Overview}
\label{sec:logic-intro}
Modular reasoning about memory in the Yarrow programming language is different 
from other programming languages, as we have to 
(1) reclaim resources used to reason about memory in regions
that are captured by continuations, and 
(2) take into account that the configurations of fibers 
and regions can change between suspending and resuming computations 
in effect handlers; an effect handler can potentially 
start and end regions, or install and uninstall other effect handlers, before using its continuation.
Our choice of \emph{logical resources} and the \emph{effectfull weakest precondition} 
are key to how we handle this complexity.
\paragraph{Logical resources} We use three types of logical resources.
The first two are points-to resources used for the stack and the heap.
\begin{align*}
  \loc \heapptr \val \qquad &\text{heap location $\loc$ stores value $\val$
  (or alternatively a closure identifier, \ie $\loc \heapptr \functionid$)} \\
  \loc \stackptr \val \qquad &\text{stack location $\loc$ stores value $\val$
  (or alternatively a closure identifier, \ie $\loc \stackptr\functionid$)}
\end{align*}
%
As we assume that the memory pointed to by heap references are freed by the garbage collector, 
YL, like most other Iris-based program logics, does not have to 
handle revocation of heap points-to resources explicitly; they can simply be dropped 
in proofs when they are no longer needed; this is always allowed in Iris since it is an \emph{affine} separation logic, \ie{} we have $\prop \ast \propB \proves \prop$, for any $\prop$ and $\propB$.
Stack points-to resources used for references allocated in regions are, however, different:
when a region ends, or a continuation captures fibers when an effect is performed, 
we need to reclaim stack points-to resources as it is unsafe to keep using them 
according to the operational semantics. 
Therefore, the proofs 
in our logic use an additional resource, $\stackrs{\fdoms}$, to keep track of the shape of the stack, \ie{} the state of the list of fibers that make up the stack.
This allows us to always know which stack points-to resources to revoke when regions end or effects are performed.
To formally describe the resource $\stackrs{\fdoms}$ we need the following additional definitions:
\begin{align*}
  \sdom \in \Domain & \eqdef{} \text{List} ~ \Loc \quad
  \sdoms \in \Domains \eqdef{} \text{List} ~ \Domain \quad
  \fdoms \in \FiberDomains \eqdef{} \text{List} ~ (\Operation \times \Domains) \\
  \stackrs{\fdoms} \quad &\text{the current stack is made of out fibers with the domains represented by $\fdoms$}
\end{align*}
The $\stackrs{\fdoms}$ resource together with stack points-to resources for all the locations in $\fdoms$ 
gives us the full picture of the fibers that make up the stack, as $\FiberDomains$ only differ from physical 
$\Fibers$ except for the values stored in the fibers' locations which can be recovered from the stack points-to resources. 
As we will see later in this section, proofs of programs that do not use
region-based memory management or effect handlers, can forget about this resource.
\paragraph{Effectful weakest precondition} To prove safety of programs in YL, 
we use an effectful weakest precondition $\ewprec{\expr}{\row}{\Phi}$, 
where the effect row $\row$ represent the effects that are handled in the expression $\expr$, and 
the postcondition $\Phi$ is a predicate on the return value of the expression $\expr$.
The type of the effect row is shown below and uses the notion of a \emph{protocol}. 
Our protocols are based on a similar definition by \citet{de2021separation}, but are redesigned to take into account the fiber domains.
\begin{align*}
  \prot \in \Protocol \eqdef{} & \Val \rightarrow \FiberDomains  \rightarrow  (\Val \rightarrow \FiberDomains \rightarrow \iProp) \rightarrow \iProp \\
  \row \in \Row \eqdef{} & \Operation \fpfn (\AffineMode \times \Protocol)
\end{align*}
Protocols work as a contract between the program performing the effect and the effect handler. 
Our protocols are parameterized by the fiber domains used by the \textlang{Stack} resource
since both the effect handler and the program performing the effect
can use $\stackrs{\fdoms}$ for revocation and we need to know the state of $\fdoms$
they are handed by the other end.
We delay explaining the details of protocols till we present the reasoning rules for performing and 
handling effects, as these are the only rules that use them.

\paragraph{The adequacy theorem}
Proofs in our logic start by assuming the resource $\stackrs{([(-,[[]])])}$ and 
no points-to resources. This reflects that when a program starts executing, we install an initial fiber 
with one initial region,
such that programs can make local allocations at the top level without having 
to explicitly start a region.
Specifically, the adequacy theorem of YL used to prove safety and functional correctness 
of programs says: 
\textit{For all expressions} $\expr$ \textit{and postcondition} $\Phi$ \textit{, if} 
$\stackrs{([(-,[[]])])} \proves \ewprec{\expr}{\row}{\Phi}$ \textit{is provable,} 
$\expr$ \textit{can always take a step in the operational semantics of \Cref{sec:semantics} 
or it} \textit{reduces to a value} $\val$ \textit{for which} $\Phi(\val)$ \textit{holds}.

\subsection{Reasoning Rules}
We divide the reasoning rules of the program logic into four categories that we proceed to go over: 
(1) structural rules (2) rules for memory management (3) rules for performing effects and
(4) rules for handling effects. In \Cref{sec:verification-example}, 
we show how the reasoning rules can be used to verify an example. 

\paragraph{Structural rules} 
We display selected structural rules of the program logic in \Cref{fig:program-logic-rules-1}.
The rules reassemble those of other Iris-based separation logics,
except we have to take into account the effect row in the effectful weakest precondition. 
For instance, in the monotonicity rules (EWP-Mono and EWP-Mono-Pers) and the frame rules (EWP-Frame and EWP-Frame-Pers),
we use the persistently modality when the effect row 
$\row$ in $\ewprec{\val}{\row}{\Phi}$ is not exclusively a row with one-shot effects; 
$\row$ lists the effects $\expr$ can use and 
if $\expr$ can perform an effect where the continuation in the effect handler 
is used multiple times (a multi-shot effect), 
then it is unsound to rely on exclusively owned resources when modifying
the postcondition, intuitively since we are modifying the
\begin{wrapfigure}{r}{0.77\textwidth}
\vspace{-5mm}
\begin{minipage}{\linewidth}
\begin{align*}
  \Phi ~ \val \proves \ewprec{\val}{\row}{\Phi} 
  & \qquad \text{(EWP-Val)} \\[0.7em]
  \onceRow{\rho} \wand (\forall \val. ~ \Phi ~ \val \wand \Psi ~ \val) \ast \ewprec{\val}{\row}{\Phi} \proves \ewprec{\val}{\row}{\Psi}
  & \qquad \text{(EWP-Mono)} \\[0.7em]
  \always(\forall \val. ~ \Phi ~ \val \wand \Psi ~ \val) \ast \ewprec{\val}{\row}{\Phi} \proves \ewprec{\val}{\row}{\Psi}
  & \qquad \text{(EWP-Mono-Pers)} \\[0.7em]
  \wprec{\expr}{v. ~ \ewprec{\nlctx[\val]}{\row}{\Phi}}\proves \ewprec{\nlctx[\expr]}{\row}{\Phi}
  & \qquad \text{(EWP-Bind)} \\[0.7em]
  \expr_1 \bstep_{\text{pure}} \expr_2 \ast \later(\ewprec{\expr_2}{\row}{\Phi}) \proves \ewprec{\expr_1}{\row}{\Phi}
  & \qquad \text{(EWP-Pure)} \\[0.7em]
  \onceRow{\rho} \wand \prop \wand \ewprec{\expr}{\row}{\Phi} \proves \ewprec{\expr}{\row}{\val. ~ \Phi ~ \val  \ast \prop}
  & \qquad \text{(EWP-Frame)} \\[0.7em]
  \always \prop \wand \ewprec{\expr}{\row}{\Phi} \proves \ewprec{\expr}{\row}{\val. ~ \Phi ~ \val  \ast \prop}
  & \qquad \text{(EWP-Frame-Pers)} \\[0.7em]
  \row' \subseteq \row \wand \ewprec{\expr}{\row'}{\Phi} \proves \ewprec{\expr}{\row}{\Phi}
  & \qquad \text{(EWP-Row1)} \\[0.7em]
  \prot' \protImpl \prot \wand \ewprec{\expr}{\rowElem{\ope}{\affineVar}{\prot'} \rowComp \row}{\Phi} \proves 
  \ewprec{\expr}{\rowElem{\ope}{\affineVar}{\prot} \rowComp \row}{\Phi}
  & \qquad \text{(EWP-Row2)} \\[0.7em]
  \infer
  {\ope_1 \neq \ope_2 \ast \ewprec{\expr}{\rowElem{\ope_2}{\affineVar_2}{\prot_2} \rowComp \rowElem{\ope_1}{\affineVar_1}{\prot_1} \rowComp \row}{\Phi}}
  {\ewprec{\expr}{\rowElem{\ope_1}{\affineVar_1}{\prot_1} \rowComp \rowElem{\ope_2}{\affineVar_2}{\prot_2} \rowComp \row}{\Phi}}
  & \qquad \text{(EWP-Row3)} 
\end{align*}
\caption{Selected structural program logic rules.}
\label{fig:program-logic-rules-1}
\end{minipage}
\end{wrapfigure} 
postcondition every time the continuation is used
to resume computation in $\expr$.
There are three structural rules involving effect rows: EWP-Row1
rows lets the user of the logic weaken the effect row
(subset inclusion on effect rows $\row_1 \subseteq \row_2$ holds if every binding in $\row_1$ occurs in $\row_2$), 
EWP-Row2 is used for weakening a protocol (ordering of protocols is defined as
$\prot_1 \protImpl \prot_2 \eqdef$ \\ $\always(\forall v ~ \fdoms ~ \Phi. ~ \prot_1 ~ v ~ \fdoms ~ \Phi \wand \prot_2 ~ v ~ \fdoms ~ \Phi)$) 
\footnote{For a
  proposition $\always P$, the persistently modality $\always$ enforces that $P$ should be proven without 
 any exclusive owned resources such as a point to resource.}, 
and the order of effects that have different operations can be changed using EWP-Row3
(~$\cdot$ is the cons operation on effect rows).
\paragraph{Rules for memory management} 
Selected rules for memory management are shown in \Cref{fig:program-logic-rules-2}.
The rules for managing the heap (EWP-Alloc-Global and EWP-Store-Global) 
are standard for Iris-based program logics.
The rules for managing fibers, and the regions we associate with stack frames in the fibers, centers around the 
$\stackrs{\fdoms}$ resource; EWP-Region inserts a new empty stack frame in the most recently installed fiber. 
A local allocation (EWP-Alloc-Local) inserts a new location, in the top most stack frame of the most recently installed fiber, 
and gives the user a stack points-to resource $\loc \stackptr \val$. The rules for manipulating local references, \eg (EWP-Local-Local), 
work just like their counterpart rules for the heap, \ie owning a points-to resource for a location $\loc$ 
implies that the location belongs to a region that has not ended yet;
what makes this possible is how deallocation of stack points-to resources work:
when a region ends in the EWP-End rule, 
one has to provide the points-to resources for all the stack reference allocated in the region, 
and then, separately, prove (without those stack points-to resources) that the postcondition 
$\Phi$ holds for a stack without the topmost region.
\begin{figure}[h]
\vspace{-3mm}
\begin{align*}
  \later(\forall \loc. ~ \loc \heapptr \val \wand \Phi ~ (\haddr{\loc})) \proves \ewprec{\ealloc{\globalMode}{\val}}{\row}{\Phi}
  & \qquad \text{(EWP-Alloc-Global)} \\[0.7em]
  \infer
  {
    \hbox{$\begin{array}{c}
    \stackrs (\fdoms \mdoubleplus [(\ope, \sdoms \mdoubleplus [\sdom])]) \asts \\
    \later (\forall \loc. ~ \stackrs (\fdoms \mdoubleplus [(\ope, \sdoms \mdoubleplus [\sdom \mdoubleplus [(\loc, \val)]])]) \ast \loc \stackptr \val \wand \Phi ~ (\saddr \loc))
    \end{array}$}
   } 
  {\ewprec{\ealloc{\localMode}{\val}}{\row}{ \Phi }}
  & \qquad \text{(EWP-Alloc-Local)} \\[0.7em]
  \loc \heapptr w \ast \later(\loc \heapptr \val \wand \Phi ~ ()) \proves \ewprec{\estore{(\haddr{\loc})}{\val}}{\row}{\Phi}
  & \qquad \text{(EWP-Store-Global)} \\[0.7em]
  \loc \stackptr \val \ast \later(\loc \stackptr \val \wand \Phi ~ \val) \proves \ewprec{\eload{(\saddr{\loc})}}{\row}{\Phi}
  & \qquad \text{(EWP-Load-Local)} \\[0.7em]
  \infer
  {\stackrs(\fdoms \mdoubleplus [(\ope, \sdoms)]) \ast 
    \later(\stackrs(\fdoms \mdoubleplus [(\ope, \sdoms \mdoubleplus [])])
    \wand \ewprec{\eend{\expr}}{\row}{\Phi})}
  {\ewprec{\eregion{\expr}}{\row}{\Phi}}
  & \qquad \text{(EWP-Region)} \\[0.7em]
  \infer
  {
    \hbox{$\begin{array}{c}
    \stackrs(\fdoms \mdoubleplus [(\ope, \sdoms \mdoubleplus [\sdom])]) \ast 
    \Sep_{(\loc, \val) \in \sdom} \loc \stackptr \val \asts 
    \later(\stackrs(\fdoms \mdoubleplus [(\ope, \sdoms)]) \wand \Phi ~ \val)
    \end{array}$}
  }
  {\ewprec{\eend{\val}}{\row}{\Phi}}
  & \qquad \text{(EWP-End)} \\[0.7em]
  \later(\forall \loc ~ \functionid. ~ \loc \heapptr \functionid \wand \Phi ~ (\vlambda{\haddr{\loc}}{\functionid}{f}{\var}{\expr})) \proves
  \ewprec{\elambda{\globalMode}{f}{\var}{\expr}}{\row}{\Phi}
  & \qquad \text{(EWP-Func-Global)} \\[0.7em]
  \infer
  {\loc \stackptr \functionid \ast \later(\loc \stackptr \functionid \wand \ewprec{\expr[(\vlambda{\saddr{\loc}}{\functionid}{f}{\var}{\expr})/f][\val/\var]}{\row}{\Phi})} 
  {\ewprec{(\vlambda{\saddr{\loc}}{\functionid}{f}{\var}{\expr}) ~ \val}{\row}{\Phi}}
  & \qquad \text{(EWP-App-Local)}
\end{align*}
\caption{Selected program logic rules for memory management.}
\label{fig:program-logic-rules-2}
\vspace{2mm}
\end{figure}
Lastly, we have allocation of functions (EWP-Func-Global) and application of functions (EWP-App-Local) which also interferes with memory management as 
function closures are explicitly allocated in our language.
Allocation of function closures is much like allocation of references: in EWP-Func-Global, the user of the logic 
obtains a heap points-to resource with a location $\loc$ pointing to an identifier $\functionid$, 
the $(\haddr{\loc}, \functionid)$ pair is then used in the $\lambda$-abstraction value 
(allocation of local closures in region happens similarly, except the fibers are updated as in EWP-Alloc-Local).
When applying a lambda abstraction, one must show that the function closure is not deallocated
by proving ownership of the points-to resource associated with the $\lambda$-abstraction value (EWP-App-Local);
for globally allocated function closures, this is easy as heap points-to resources are persistent, 
but stack points-to resource are deallocated when regions end as seen in the EWP-End rule. 
\paragraph{Rules for performing effects}
Suppose we perform an effect with operation $\ope$ as in the proof goal of the EWP-Do-Once rule in \Cref{fig:program-logic-rules-3}, 
\ie we want to prove $\ewprec{\edo{\ope}{\val}}{\rowElem{\ope}{\onceMode}{\prot} \rowComp \row}{\Phi}$. 
Here, the effect row tells us that the effect we are performing is a one-shot effect, and that 
the contract we have with the effect handler is described by the protocol $\Psi$.
First, we have to provide the stack resource 
$\stackrs{(\fdoms_1 \mdoubleplus [(\ope, \sdoms)] \mdoubleplus \fdoms_2)}$
together with stack points-to resources for a map $m$: 
$ \Sep_{(\loc, \val) \in m} \loc \stackptr \val$.
The predicate $\compatible ~ \ope ~ \sdoms ~ \fdoms_2 ~ m$ asserts that
$\fdoms_2$ does not contain a fiber with operation $\ope$, and 
the references in $m$ are equal to those allocated in the fibers ($[(op, \sdoms)] \mdoubleplus \fdoms_2$).
Intuitively, this means we have found the fiber of the effect we are performing, and at the same time, 
deallocated all the references that the continuation of the effect handler captures in its regions.
The last obligation of the EWP-Do-Once rule, is to satisfy the protocol: 
\begin{align*}
\prot ~ \val ~ \fdoms_1 ~ (\lambda ~ w ~ \fdoms_1'. ~ \stackrs{(\fdoms_1' \mdoubleplus [(\ope, \sdoms)] \mdoubleplus \fdoms_2)}
\wand \Sep_{(\loc, \val) \in m} \loc \stackptr \val \wand \Phi ~ w) 
\end{align*}
We always instantiate a protocol with the argument value $\val$ provided when performing the effect, 
the part of the stack $\fdoms_1$ that the effect handler is executing with, and a proof obligation 
that must be proven when resuming computation with a continuation using return value $w$ and 
a potentially updated stack described by $\fdoms_1'$ 
(the stack can change if the effect handler starts or end regions and installs or uninstall 
other effect handlers).
The proof obligation says that given the stack resource where the fibers 
that the continuation captured are back onto the stack $\stackrs{(\fdoms_1' \mdoubleplus [(\ope, \sdoms)] \mdoubleplus \fdoms_2)}$, 
and ownership of the stack points-to resources for the references in the regions of these fibers $\Sep_{(\loc, \val) \in m} \loc \stackptr \val$, 
the post-condition of the proof goal we had when performing the effect $\Phi ~ w$ must hold.
The only difference between the EWP-Do-Once rule for one-shot effects and the EWP-Do-Many rule for multi-shot effects 
is that when proving the post-condition of the proof goal we had when performing the effect $\Phi ~ w$, 
we can not assume stack points-to resources for the references in the regions of the fibers that the continuation 
capture.
\newcommand{\protaff}{\mathit{protAff}}
\begin{figure}[h]
\vspace{-3mm}
\begin{align*}
  \infer
  {
    \hbox{$\begin{array}{c}
      \exists \fdoms_1 ~ \fdoms_2 ~ \sdoms ~ m. ~ \stackrs{(\fdoms_1 \mdoubleplus [(\ope, \sdoms)] \mdoubleplus \fdoms_2)} \ast
      \Sep_{(\loc, \val) \in m} \loc \stackptr \val \ast
      \compatible ~ \ope ~ \sdoms ~ \fdoms_2 ~ m \vspace{1.5mm}\\
      \asts \later \prot ~ \val ~ \fdoms_1 ~ (\lambda ~ w ~ \fdoms_1'. ~ \stackrs{(\fdoms_1' \mdoubleplus [(\ope, \sdoms)] \mdoubleplus \fdoms_2)} \wand
      \Sep_{(\loc, \val) \in m} \loc \stackptr \val \wand \Phi ~ w) 
    \end{array}$} 
  }
  {\ewprec{\edo{\ope}{\val}}{\rowElem{\ope}{\onceMode}{\prot} \rowComp \row}{\Phi}}
  \qquad \text{(EWP-Do-Once)} & \\[0.7em]
  \infer
  {
    \hbox{$\begin{array}{c}
      \exists \fdoms_1 ~ \fdoms_2 ~ \sdoms ~ m. ~ \stackrs{(\fdoms_1 \mdoubleplus [(\ope, \sdoms)] \mdoubleplus \fdoms_2)} \ast
      \Sep_{(\loc, \val) \in m} \loc \stackptr \val \ast
      \compatible ~ \ope ~ \fdoms_1 ~ [(\ope, \sdoms)] ~ \fdoms_2 ~ m \vspace{1.5mm}\\
      \asts \later \prot ~ \val ~ \fdoms_1 ~ (\lambda ~ w ~ \fdoms_1'. ~ \stackrs{(\fdoms_1' \mdoubleplus [(\ope, \sdoms)] \mdoubleplus \fdoms_2)} \wand \Phi ~ w) 
    \end{array}$}
  }
  {\ewprec{\edo{\ope}{\val}}{\rowElem{\ope}{\manyMode}{\prot} \rowComp \row}{\Phi}}
  \qquad \text{(EWP-Do-Many)} & \\[0.7em]
    \infer
  {
    \hbox{$\begin{array}{c}
    \protaff ~ \affineVar ~ \prot \ast \stackrs{\fdoms} \ast \later \hdlglb ~ \prot ~ \Phi_e ~ \var ~ k ~ h ~ y ~ r ~ \rho ~ \mathit{m\rho} ~ \Phi \asts \\
    \later \big(\stackrs{(\fdoms \mdoubleplus [(\ope, [])])} \wand \ewprec{\expr}{\rowElem{\ope}{\affineVar}{\prot} \rowComp \row}
      {\val. ~ \exists \fdoms'. ~ \Phi_e ~ \fdoms' ~ \val \ast \stackrs{(\fdoms' \mdoubleplus [(\ope, [])])} }\big)
    \end{array}$}
  }
  {\ewprec{(\ehandle{\globalMode}{\affineVar}{\expr}{\ope}{\var}{\con}{h}{y}{r})}{\row}{\Phi}}
  \qquad \text{(EWP-Try)} &
\end{align*}
\caption{Selected program logic rules performing and handling effects.}
\label{fig:program-logic-rules-3}
\end{figure}
\paragraph{Rules for handling effects}
To install an effect handler with allocation of the continuation closure on the heap, 
we use the EWP-Try rule in \Cref{fig:program-logic-rules-3} for both one-shot and multi-shot effects.
In this rule, we must first prove that the affinity of the protocol $\Psi$ matches the affinity variable $\affineVar$
(we return to how this is done when we present an example protocol in \Cref{sec:verification-example})
and provide the stack resource $\stackrs{\fdoms}$.
There is then two proof obligations: (1) prove the handler branch and the return branch are safe to execute; this is captured 
by the $\hdlglb$ predicate whose definition we return to shortly, 
and (2) prove that the expression $\expr$ for which the effect is installed is safe to execute; this is captured by 
the following proof goal: 
\begin{align*}
  \stackrs{(\fdoms \mdoubleplus [(\ope, [])])} \wand \ewprec{\expr}{\rowElem{\ope}{\affineVar}{\prot} \rowComp \row}
      {\val. ~ \exists \fdoms'. ~ \Phi_e ~ \fdoms' ~ \val \ast \stackrs{(\fdoms' \mdoubleplus [(\ope, [])])}}
\end{align*}
Here we get to assume the stack resource
$\stackrs{(\fdoms \mdoubleplus [(\ope, [])])}$ with a fiber installed for the
new effect. Moreover, the new effect is added to the effect row in the effectful
weakest precondition with affinity $\affineVar$ and protocol $\Psi$ (it is here
we see that the protocol is picked as part of the EWP-Try rule and then used as
a contract for performing the effect in $\expr$, because the protocol $\Psi$ is
added to the effect row in the effectful weakest precondition). As 
postcondition we must prove a user-defined postcondition $\Phi_e$, which we
instantiate with the return value $\val$ and the updated fiber domains
$\fdoms$'. Moreover, we must also prove that the current fibers can be described by the
stack resource $\stackrs{(\fdoms' \mdoubleplus [(\ope, [])])}$ which captures that the
top most fiber is the one for the effect with operation $\ope$, but that all the
fibers before that are now $\fdoms'$. The reason why the fibers installed before
the effect with operation $\ope$ can change from $\fdoms$ to $\fdoms'$, is that
inside $\expr$, effects in the effect row $\row$ may be performed, and the
handlers for these effects can install other effects and, perhaps more likely, make
changes to regions by \eg allocating local references.

Next, we turn our attention to the $\hdlglb$ predicate which captures the 
correctness of the return branch $r$ and the handler branch $h$. We show its definition in \Cref{fig:handler-global}.
\begin{figure}[h]
\vspace{-3mm}
\begin{align*}
&\hdlglb ~ \prot ~ \Phi_e ~ \var ~ k ~ h ~ y ~ r ~ \rho ~ \mathit{m\rho} ~ \Phi \eqdef{}
 \mathit{m\rho} = \onceMode \rightarrow \onceRow ~ \rho ~ \land \condAlways{\mathit{m\rho}} ~ \big(\\
 &\quad (* ~ \mathit{Return} ~ \mathit{branch:} ~ *)\\
 &\quad (\forall \val ~ \fdoms. ~ \Phi_e ~ \fdoms ~ \val \wand \stackrs ~ \fdoms ~
    \wand \ewprec{r[\val/y]}{\row}{\Phi}) ~ \land \\
&\quad (* ~ \mathit{Handler} ~ \mathit{branch:} ~ *) \\
 &\quad (\forall \val ~ k'. ~ (\exists \fdoms. ~ \stackrs ~ \fdoms \asts 
  \prot ~ \val ~ \fdoms ~ (\lambda w ~ \fdoms'. ~ \forall \Phi'. ~ \stackrs ~ \fdoms' \wand 
  \later \hdlglb ~ \prot ~ \Phi_e ~ \var ~ k ~ h ~ y ~ r ~ \rho ~ \mathit{m\rho} ~ \Phi' \wand
  \ewprec{k' ~ w}{\rho}{\Phi'})) \\
 &\qquad\qquad\hspace{1mm} \wand \ewprec{h[k'/k][\val/x]}{\row}{\Phi})\big)
\end{align*}
\caption{Definition of the predicate $\hdlglb$.}
\label{fig:handler-global}
\vspace{1mm}
\end{figure}
This predicate is divided into two parts, one part for the return branch and one part for the handler branch.
Around the two branches, there is a persistently modality conditional on the affinity mode $\mathit{m\rho}$; 
when $\mathit{m\rho}$ is $\onceMode$, we can remove the persistently modality and use exclusively owned resources to 
prove correctness of the return and handler branche. 
When the persistently modality is removed, it is required that the existing effect row $\rho$ can not have any multi-shot effects.
This is because the program $\expr$ that we install the effect for can also use effects from $\rho$. 
If $\expr$ uses a multi-shot effect, then execution can enter the handler branch $h$ and return branch $r$ 
many times without using a continuation (we reason about the use of continuations inside the proof of the handler branch), 
thus we can not rely on exclusively owned resources when reasoning about $h$ and $r$.

In the proof of the return branch $r$, we get to assume the postcondition $\Phi_e ~ \fdoms ~ \val$
of the expression $\expr$ that the effect was installed for, where $\val$ is the return value of $\expr$ and 
$\fdoms$ corresponds to the fiber domains when the execution of $\expr$ ended, 
together with the  stack resource $Stack ~ \fdoms$. 
In the proof of the handler branch $h$, $k'$ is the continuation which, together 
with the argument $\val$ that was used to perform the effect, 
is substituted for the binders $k$ and $x$. We can assume the stack resource $Stack ~ \fdoms$, and 
the protocol instantiated as below:
\begin{align*}
  \prot ~ \val ~ \fdoms ~ (\lambda w ~ \fdoms'. ~ \forall \Phi'. ~ \stackrs ~ \fdoms' \wand
  \later \hdlglb ~ \affineVar ~ \prot ~ \Phi_e ~ \var ~ k ~ h ~ y ~ r ~ \rho ~ \mathit{m\rho} ~ \Phi' \wand
  \ewprec{k' ~ w}{\rho}{\Phi'})
\end{align*}
Here we see the opposite end of the contract that a protocol $\prot$ establishes 
between an effect handler and the program performing the effect
(we saw the other end when we discussed the EWP-Do-Once and EWP-DO-Many rules).
Note that in the handler branch we get to assume the protocol above, it is not a proof obligation.
The predicate on the value $w$ and the fiber configuration $\fdoms'$ 
that we instantiate the protocol with on this end,
says that to get a proof of the effectful weakest precondition 
for the continuation $k'$ that we can use in the proof of the handler branch, 
we must provide the stack resource $\stackrs ~ \fdoms'$ and prove 
the recursive occurrence of the $\hdlglb$ predicate. 
The reason we have to prove the recursive occurrence of the $\hdlglb$ 
predicate is that the effect handler is reinstalled as a part of the continuation.

In the last few paragraphs, we went over the proof rules for installing an effect handler 
with the global mode $\globalMode$. If we want to use the local mode $\localMode$, 
and install the continuation closure in the region that the effect handler uses,
we need to use a version of EWP-Try with the only modification that $\hdlglb$
is instead $\hdlloc$. We have included the definition of $\hdlloc$ in \Cref{sec:cont-rule}; 
it works like $\hdlglb$, except each time the continuation is used in the handler branch, 
one must prove that the continuation closure is still alive by showing ownership 
of a stack points resource (the resource is provided as part of $\hdlloc$) 
with the closure identifier that the continuation uses
--- like when applying a function in the EWP-App-Local rule. 

\subsection{Verification of the Increment First Example}
\label{sec:verification-example}
In this subsection, we go over a proof sketch of the Increment First Example in \Cref{fig:inc-fst-impl}
with a focus on using the rules for performing and handling effects. 
Our goal is to prove $\stackrs{([(-,[[]])])} \proves \ewprec{\textlang{handle\_example2 ()}}{\emptyRow}{\TRUE}$,
so we can use the adequacy theorem of YL from \Cref{sec:logic-intro}. 
We skip forward a bit to line 3 in the code just before the
\langkw{try}-construct. At this point in the proof, we have the resources
$\stackrs{([(-,[[]; []; [\textlang{x}]])])}$ and $\textlang{x} \stackptr 0$
reflecting that the regions of \textlang{handle\_example2} and \textlang{example2} 
are started, and in the region started in \textlang{example2}, we have allocated 
the local reference \textlang{x} on line 2.
As we need to reason about an effect handler, we use the EWP-Try rule.
Using this rule, we must pick a protocol that captures the interaction between 
the effect handler and \textlang{example2 x} which is the code the effect is installed
for.
Throughout this paper, we use a particular type of protocol, 
namely the \textbf{\textit{send-receive protocol}}, an extension of the \textit{send-receive protocol} from \citet{de2021separation} 
that also takes into account the fiber domains used with the \textlang{Stack} resource:
\begin{align*}
  \stackProt{\vec{x}}{\val_1, \fdoms_1}{P}{\affineVar}{\vec{y}}{\val_2, \fdoms_2}{Q} \eqdef 
  \lambda v ~ \fdoms ~ \Phi. ~ \exists \vec{x}. ~ \val = \val_1 \land \fdoms = \fdoms_1 \ast P ~ x \ast
  \condAlways{\affineVar}(\forall \vec{y}. ~ Q \wand \Phi ~ \val_2 ~ \fdoms_2).%
\end{align*}%
When a send-receive protocol on the form above is used, 
it means that the value argument $\val$ must be equal 
to $\val_1$ and that the fiber domains, when the fibers that the continuation capture are removed from the stack,
must be equal to $\fdoms_1$. The proposition $P$ works as a precondition that the proof of a program performing the effect must prove, 
and the effect handler can assume when handling the effect.
When the effect handler calls the continuation, it must call it with an 
argument equal to $\val_2$, be in a state where the fiber domains is equal to $\fdoms_2$ 
($\fdoms_2$ are the fiber domains before the fibers of the continuation are appended to the stack) 
and prove that $Q$ holds. Thus, $P$ and $Q$ can be seen as precondition and postcondition of $\edo{\ope}{\val}$. 
To make the send-receive protocol expressive, it existentially quantifiers over the list of variables $\vec{x}$ 
in $\val_1$, $\fdoms_1$, $P$, $\val_2$, $\fdoms_2$ and $Q$, 
and universally quantifiers over $\vec{y}$ in $\val_2$, $\fdoms_2$ and $Q$.
The affinity variable $\affineVar$ controls whether the send-receive protocol 
is for a one-shot or multi-shot effect; to control this we use the \emph{persistently modality} $\always$. 
When the affinity variable $\affineVar$ is \onceMode, 
$\condAlways{\affineVar} ~ P$ is the same as $\always P$, 
whereas if $\affineVar$ is \manyMode, the persistently modality is removed and 
$\condAlways{\affineVar} ~ P$ becomes $P$. In essence, the persistently modality is conditional on $\affineVar$.
The send-receive protocol, together with the composition protocol below that composes two existing protocols, 
are the only protocols we use in this paper: 
\begin{align*}
  \prot_1 \protComp \prot_2 \eqdef \lambda v ~ \fdoms ~ \Phi. ~ \prot_1 ~ v ~ \fdoms ~ \Phi \lor \prot_2 ~ v ~ \fdoms ~ \Phi
\end{align*}
Here is the concrete send-receive protocol we use in this proof with the EWP-Try rule:
\begin{align*}
  INC \eqdef ~
  \stackProt{\fdoms ~ \loc_1 ~ \loc_2 ~ \vInt ~}{(\saddr{\loc_1}, \saddr{\loc_2}), \fdoms}{\loc_1 \stackptr \vInt}{\onceMode}{}{(), \fdoms}{\loc_1 \stackptr (\vInt + 1)}
\end{align*}
The $INC$ protocols says that the effect can be called with a pair of stack addresses $(\saddr{\loc_1}, \saddr{\loc_2})$ as argument 
and any fiber domains $\fdoms$ in the \textlang{Stack} resource. 
For the first entry in the pair, the stack points-to resource $\loc_1 \stackptr \vInt$ is required 
as precondition. When the effect resumes, it resumes with the unit value and the fiber domains unchanged. 
The postcondition gives back the stack points-to resource but the integer it points to is now incremented: $\loc_1 \stackptr  (\vInt + 1)$.

Now that we have picked our protocol, we can proceed to prove the effect handler 
on line 3 in \Cref{fig:inc-fst-impl} and the code \textlang{example2 x} that the effect is installed for,
as required by the EWP-Try rule. 
When a protocol $\Psi$ is made using send-receive and composition protocols, showing the $\protaff ~ \affineVar ~ \prot$ obligation 
in the EWP-Try rule amounts to checking whether the affinity variable used in send-receive protocols matches $\affineVar$.
\footnote{In the Rocq formalization, we use \emph{upward closures} of protocols as
in \citet{VilhenaPhesis}, but the protocols used in this paper are equivalent 
to their upward closures when the $\protaff$ property holds.}
In this example, we are installing the effect handler on line 3 with the $\onceMode$ mode which matches the mode used in 
the $INC$ protocol.

\paragraph{Proof of \textlang{example2}}
The \textlang{example2} function is used with the argument \textlang{x} for which we have the stack points-to resource 
$\textlang{x} \stackptr 0$.
Because $\stackrs{([(-,[[]; []; [\textlang{x}]])])}$ is the state of the stack resource prior to using EWP-Try, 
we can also assume the stack resource $\stackrs{([(-,[[]; []; [\textlang{x}]]); (\textlang{IncFst}, [])])}$,
updated with a new fiber for the effect handler that we just installed.
We pick $\Phi_\expr \eqdef \lambda \fdoms ~ \_. ~ \fdoms = [(-,[[]; []; [\textlang{x}]])] \ast \textlang{x} \stackptr 1$ 
in EWP-Try which makes our proof obligation this:
\begin{align*}
\ewprec{\textlang{example2 x}}
{\rowElem{\textlang{IncFst}}{\onceMode}{~ INC}}
{\stackrs{([(-,[[]; []; [\textlang{x}]]);(\textlang{IncFst}, [])])} \ast \textlang{x} \stackptr 1}
\end{align*}
Let us skip ahead in the proof to where the effect is performed. Here, 
we have the resources $\stackrs{([(-,[[]; []; [\textlang{x}]]);(\textlang{IncFst}, [[y]])])}$, 
$\textlang{x} \stackptr 0$ and $\textlang{y} \stackptr 1$, 
which reflects that the region of \textlang{example2} started and the reference \textlang{y} is allocated in it.
The proof obligation at this point is:
\begin{align*}
&\text{ewp}~ \textlang{\textbf{do} IncFst (x, y)}_{(\textlang{IncFst} ~ : ~ (\onceMode, ~ INC))} 
\big\{\val. ~
  \text{ewp} ~ \textlang{\textbf{end} (} \val \textlang{;\textbf{assert} (!x + !y = 2))}
  _{(\textlang{IncFst} ~ : ~ (\onceMode, ~ INC))} \\ & \quad
  \{\stackrs{([(-,[[]; []; [\textlang{x}]]);(\textlang{IncFst}, [])])} \ast \textlang{x} \stackptr 0\}
\big\}
\end{align*}
Because the effect we want to perform is a one-shot effect, as seen in the effect row of our proof obligation, 
we use the EWP-Once rule. This rule says that we must provide the stack points-to resources for all
locations of the regions in the fibers up to and including the fiber for the effect we perform, as these are the fibers
captured by the continuation. In our case, this amounts to the fiber $(\textlang{IncFst}, [[\textlang{y}]])$
with one region containing \textlang{y} which we provide the stack points-to resource 
$\textlang{y} \stackptr 1$ for. 
The fiber $[(-,[[]; []; [\textlang{x}]])$ prior 
to the $(\textlang{IncFst}, [[\textlang{y}]])$ fiber, is what the effect handler gets to use, 
but the protocol $INC$ makes no assumption about the contents of this fiber.
As we are using \textlang{(x, y)} as argument 
to the effect, the protocol $INC$ says that we must provide $\textlang{x} \stackptr 0$ to the effect handler.
The postcondition of the protocol gives us back the stack points-to resource incremented ($\textlang{x} \stackptr 1$), 
and the fiber prior to the $\textlang{IncFst}$ fiber is left unchanged. 
When proving the postcondition of our proof obligation, we also get to assume 
$\stackrs{([(-,[[]; []; [\textlang{x}]]); (\textlang{IncFst}, [[\textlang{y}]])])}$; 
because we were performing a one-shot effect, 
$\textlang{y}$ is still in region of the top most fiber, and we get back the points-to resource 
$\textlang{y} \stackptr 1$.
With these resources in hand, we can prove the postcondition of the proof obligation 
we listed above by satisfying the assertion and ending the top most regions with \textlang{y} in it, 
which consumes the $\textlang{y} \stackptr 1$ resource and leaves us with the 
$\stackrs{([(-,[[]; []; [\textlang{x}]]); (\textlang{IncFst}, [])])} \ast \textlang{x} \stackptr 1$ 
resources as needed.
\paragraph{Proof of the effect handler} In the proof obligation about the effect handler 
in the EWP-Try rule, we instantiate the $\hdlglb$
predicate which amounts to two proof obligations, 
one for the handler branch and one for the return branch. The obligation for the return branch 
is shown below and follows from the EWP-End rule:
\begin{align*}
  \stackrs{[(-,[[]; []; [\textlang{x}]])]} \ast \textlang{x} \stackptr 1 \proves
  \ewprec{\langkw{end} ~ (\langkw{end} ~ \val)}
  {\emptyRow}
  {\TRUE}
\end{align*}
The resources $Stack [(-, [[]; []; [x]])]$ and $\textlang{x} \stackptr 1$ are what $\textlang{example\_2 x}$ 
that we installed the effect for
return. Thus the above goal completes the proof obligations about the return branch and 
the code surrounding the effect handler which amounts to ending the two regions started in 
\textlang{handle\_inc\_fst} and \textlang{handle\_example2}. 
For the handler branch, the proof obligation is:
\begin{align*}
  \stackrs{\fdoms} \ast \textlang{x} \stackptr 0 \proves
  \ewprec{\textlang{((x, y).1 <- !(x, y).1 + 1); k ()}}
  {\emptyRow}
  {\ewprec{\langkw{end} ~ (\langkw{end} ~ \val)}
  {\emptyRow}
  {\TRUE}}
\end{align*}
It is the $INC$ protocol that dictates we get to assume $\textlang{x} \stackptr 0$ and 
$\stackrs{\fdoms}$ for some fiber domains $\fdoms$.
Further, the protocol gives us the specification for calling the continuation in 
the $\hdlglb$ predicate, which says that if we can prove the recursive occurrence of the 
$\hdlglb$ predicate, we get to assume: 
\begin{align*}
  \stackrs{\fdoms} \ast \textlang{x} \stackptr 1 \proves
  \ewprec{\textlang{k ()}}
  {\emptyRow}
  {\ewprec{\langkw{end} ~ (\langkw{end} ~ \val)}
  {\emptyRow}
  {\TRUE}}
\end{align*}
The specification captures that calling the continuation with the effect handler reinstalled 
(the effect handler is reinstalled as part of \textlang{k}),
the postcondition for the return branch that we already proved holds
(the return branch is used because the effect handler is reinstalled), 
as long as we can prove the precondition of the continuation 
$\stackrs{\fdoms} \ast \textlang{x} \stackptr 1 \proves$ as specified 
by the $INC$ protocol.
The proof obligation for the handler branch follows from the specification of the continuation above, 
as the increment of \textlang{x} turns $\textlang{x} \stackptr 0$ into $\textlang{x} \stackptr 1$, 
leaving us with the exact resources and proof goal needed. 
The only missing obligation is to prove the recursive occurrence of the $\hdlglb$ predicate, 
but this follow from Löb induction ($(\later P \Rightarrow P) \proves P$) applied at the beginning of the proof of the effect handler.


\section{Case Study: LIFO data structure}
\label{sec:stack}

In this section, we show how an effect-handler implementation of a data structure
can be made with local allocations.
We focus on a LIFO data structure, \ie a stack, 
but the approach generalizes
to other data structure implementations. 
The code of the LIFO data structure implementation is shown in \Cref{fig:stack-impl}, 
where \textlang{handle\_lifo} is the implementation of the data structure.
Our approach to implementing data structures 
using local allocations in regions is to make all allocations 
in the region of the effect handler; this also 
includes the continuations' closure.
In the implementation of the \textlang{LIFO} effect, 
we use a local reference \textlang{r} that always points-to the element
that was last inserted. The LIFO data structure is implemented as a linked-list 
of local references and has two operations, an \textlang{Insert} and a 
\textlang{Remove} operation defined using sum types. 
To handle the \textlang{Insert} operation with argument \textlang{x}
(\textlang{x} is the value we want to insert into the data structure), 
we load the head of the list from the \textlang{r} reference, 
and then update \textlang{r} with an optional of a newly allocated local reference; 
the reference points to a pair 
of the value \textlang{x} and the previous head of the list \textlang{hd}.
The protocol for the \textlang{LIFO} effect that we define shortly 
makes sure that we never try to use the \textlang{Insert} operation
on the empty data structure, hence we never assert \textlang{false} on line 8. 
Thus, when the \textlang{Remove} operation is used, we can 
update \textlang{r} by setting it to the next element in the linked-list that makes up the LIFO data structure. 
Below is the \textlang{LIFO} effect protocol that we have proven the 
implementation against: 
\begin{align*}
  \mathit{LIFO} ~ \ope ~ \sdoms ~ \mathit{lifo} \eqdef ~ & \hspace{2.5mm} 
  \stackProt{\fdoms ~ \sdom_1 ~ x ~ xs ~}{\textlang{Insert} ~ x, \fdoms \mdoubleplus [(\ope, \sdoms \mdoubleplus [\sdom_1])]}{lifo ~ xs ~ \sdom_1}{\onceMode\\&\hspace{2.5mm}}{\sdom_2 ~}{(), \fdoms \mdoubleplus [(\ope, \sdoms \mdoubleplus [\sdom_2])]}{lifo ~ (x :: xs) ~ \sdom_2} \\
  &\protComp ~ \stackProt{\fdoms ~ \sdom_1 ~ x ~ xs ~}{\textlang{Remove}, \fdoms \mdoubleplus [(\ope, \sdoms \mdoubleplus [\sdom_1])]}{lifo ~ (x :: xs) ~ \sdom_1}{\onceMode\\&\hspace{2.5mm}}{\sdom_2 ~}{x, \fdoms \mdoubleplus [(\ope, \sdoms \mdoubleplus [\sdom_2])]}{lifo ~ xs ~ \sdoms_2}
\end{align*}
Central to the specification is a predicate $\mathit{lifo} ~ xs ~ \sdom$
that keeps track of the contents of the data structure using a list 
$xs$, and a stack domain $\sdom$ which corresponds to the region 
of the effect handler.\footnote{In the Rocq formalization, we have a lemma to exchange 
between $\mathit{lifo} ~ xs ~ \sdom$ and stack points-to resources
that correspond to the references in $\sdom$, 
these resources can be used to prove the $\compatible$ predicate 
in the EWP-DO rules when performing other effects.
Similar lemmas exist for the definitions used in the protocols
of the other case studies that mention stack domains or fiber domains.} 
The list $xs$ in the predicate changes 
as one would expect from using the \textlang{Insert} and 
\textlang{Remove} operations. The stack domain $\sdom$ also changes, 
during the operation, from $\sdom_1$ when performing the \textlang{LIFO} effect 
to some other stack domain $\sdom_2$ when resuming computation.
We see that $\sdom_1$ does in fact correspond to the top most stack frame 
when performing the effect, 
as the protocol asserts that the fiber domains in which the effect handler gets 
to execute are $\fdoms \mdoubleplus [(\ope, \sdoms \mdoubleplus [\sdom_1])]$
(remember these are the fiber domains after the fibers of continuation has been captured). 
Likewise, the effect handler's stack frame has changed to $\sdom_2$ during the execution 
of the effect handler, as the protocol states that the 
fiber domains are $\fdoms \mdoubleplus [(\ope, \sdoms \mdoubleplus [\sdom_1])]$ when resuming computation
(this is the state before the fibers that the continuation capture are reinserted).
This models that the implementation of the specification 
can use the region associated with the top most stack frame
to make local allocations during the execution of the operations that the effect exposes; 
the fiber for the operation $\ope$ is some arbitrary fiber where the \textlang{LIFO} effect 
handler is installed.
\begin{figure}[h]
\vspace{-2mm}
\begin{lstlisting}
let handle_lifo f = region (
  let r = ref local None in
  try (local, once) f () with
  | effect LIFO arg k -> 
    (match arg with 
     | Insert x -> let hd = !r in r <- (Some (ref local (x, hd))); k () 
     | Remove -> (match !r with | None -> assert false | Some r -> let p = !r in r <- p.2; k p.1))
  | ret x -> x)

let example3 () = region (
  do LIFO (Insert 1); do LIFO (Insert 2); do LIFO (Insert 3); 
  assert (do LIFO Remove = 3); assert (do LIFO Remove = 2); assert (do LIFO Remove = 1))

let handle_example3 () = region (handle_lifo example3)
\end{lstlisting}
\caption{LIFO data structure implementation.}
\label{fig:stack-impl}
\vspace{-1mm}
\end{figure}


\section{Case study: Checkpointing}
\label{sec:checkpoint}
In this section, we use an effect handler to implemented 
locally allocated checkpoints; checkpoints are continuations 
that we store in the region where the effect handler is installed together with the closure of the continuation.
The implementation of the \textlang{Checkpoint} effect is shown in
\Cref{fig:checkpoint-impl}, and it is inspired by a similar example by \citet{Muhcu25}.
The \textlang{Checkpoint} effect has two operations: a \textlang{Save} operation to save a checkpoint 
at the current line in the code, and a \textlang{Retry} operation to resume computation from the last saved 
checkpoint.
In the definition of \textlang{example4} on line 7, we see how a checkpoint is saved. 
The branching on line 9, based on the value of the local reference \textlang{r} allocated in the region 
of \textlang{handle\_example}, goes to the \textlang{if}-branch as \textlang{r} points to \textlang{0} initially. 
In the \textlang{if}-branch, \textlang{r} is incremented whereafter we retry computation from the checkpoint
that was saved on line 8 such we never have to assert \textlang{false}. 
Now that \textlang{r} is no longer \textlang{0}, we go to the else branch where the assertion holds. 
The implementation of the \textlang{Checkpoint} effect on line 3 in \Cref{fig:checkpoint-impl} 
uses a local reference to store checkpoints in form of continuations. 
Continuations are locally allocated in the region of the \textlang{handle\_checkpoint} function. 
\textlang{Checkpoint} is a multi-shot effect, 
because when a checkpoint is saved, we store the continuation, such that it can be called 
later when the retry operation of the effect is performed, but we also 
use the continuation to resume computation immediately. 

Below, we specify the \textlang{Checkpoint} effect with a protocol:
\begin{align*}
  CH&ECKPOINT ~ \ope ~ \sdoms ~ \mathit{cp} ~ \mathit{noCp} ~ P_1 ~ P_2 \eqdef \\
  & \stackProt{\fdoms_1 ~ \sdom_1}{\textlang{Save}, \fdoms_1 \mdoubleplus [(\ope, \sdoms \mdoubleplus [\sdom_1])]}{(\mathit{noCp} ~ \sdom_1 \lor \mathit{cp} ~ \sdom_1) \ast P_1 ~ (\fdoms_1 \mdoubleplus [(\ope, \sdoms)])}{\manyMode\\&}
    {\fdoms_2 ~ \sdom_2 ~}{(), \fdoms_2 \mdoubleplus [(\ope, \sdoms \mdoubleplus [\sdom_2])]}{\mathit{cp} ~ \sdom_2 \ast (P_1 ~ (\fdoms_2 \mdoubleplus [(\ope, \sdoms)]) \lor P_2 ~ (\fdoms_2 \mdoubleplus [(\ope, \sdoms)]))} \\
  \protComp ~ & \stackProt{\fdoms ~ \sdom ~}{\textlang{Retry}, \fdoms \mdoubleplus [(\ope, \sdoms \mdoubleplus [\sdom])]}{\mathit{cp} ~ \sdom_1 \ast P_2 ~ (\fdoms \mdoubleplus [(\ope, \sdoms)])}{\manyMode}
    {}{(), []}{\FALSE}
\end{align*}
The protocol uses a number of predicates: 
$\mathit{cp}$ and $\mathit{noCp}$ are defined as part of the proof of the effect handler, whereas 
$P_1$ and $P_2$, which we will refer to as user defined, are defined as part of the proof of the program that uses the effect
(\cf the proof structure in \Cref{sec:verification-example}). 
Ownership of $\mathit{cp} ~ \sdom$ signifies that a checkpoint is installed,
and $\mathit{noCp} ~ \sdom$ that no checkpoint is installed; these predicates are parametrized over the stack frame
of the \textlang{Checkpoint} effect handler as this contains the region used to store checkpoints. 
To save a checkpoint, we must provide either $\mathit{noCp} ~ \sdom_1$ or $\mathit{cp} ~ \sdom_1$ as precondition together 
with the user defined predicate $P_1 ~ (\fdoms_1 \mdoubleplus [(\ope, \sdoms)])$
(the user-defined predicates are parameterized over all stack frames, except the effect handler's, 
to allow the user to make assertions about stack points-to resources in regions).
In the postcondition of the \textlang{Save} operation, 
we can assert that our new checkpoint is installed in form of ownership of the 
predicate $\mathit{cp} ~ \sdom_2$ for a new stack frame $\sdom_2$.
We also gain $P_1 ~ (\fdoms_2 \mdoubleplus [(\ope, \sdoms)])$ or $P_2 ~ (\fdoms_2 \mdoubleplus [(\ope, \sdoms)])$.
$P_1 ~ (\fdoms_2 \mdoubleplus [(\ope, \sdoms)])$ is returned when computation 
is resumed just after saving the checkpoint, but we also have to show that 
we can resume the computation using the continuation of the \textlang{Save} operation
when done through the \textlang{Retry} operation; 
in the protocol above, the specification for the \textlang{Retry} operation states that 
the predicate $P_2$ must be provided in the precondition. As computation is never resumed 
using the continuation of a \textlang{Retry} operation, the postcondition is simply false. 
Instead, computation is resumed using the saved checkpoint and ownership of 
$P_2 ~ (\fdoms_2 \mdoubleplus [(\ope, \sdoms)])$ as shown in the postcondition of the 
\textlang{Save} operation. When proving safety of \textlang{example4} in \Cref{fig:checkpoint-impl}, 
$P_1$ is instantiated with the stack points-to resource of \textlang{r} pointing to $0$ ($r \stackptr 0$), and 
$P_2$ with the resource pointing to $1$ ($r \stackptr 1$).
\begin{figure}[h]
\vspace{-2.5mm}
\begin{lstlisting}
let handle_checkpoint f = region (
  let r = ref local () in
  try (local, many) f () with
  | effect Checkpoint arg k -> (match arg with | Save -> r <- k; k () | Retry -> (!r) ())
  | ret x -> x)

let example4 r () = region (
  do Checkpoint Save; 
  if (!r = 0) then (r <- (!r + 1); do Checkpoint Retry; assert false) else (assert (!r = 1)))

let handle_example4 () = region (let r = ref local 0 in handle_checkpoint (example4 r))
\end{lstlisting}
\caption{Checkpoint Implementation.}
\label{fig:checkpoint-impl}
\vspace{-2mm}
\end{figure}


\section{Case Study: Asynchronous Computation}
\label{sec:async}
We have taken the asynchronous computation implementation by \citet{de2021separation}, 
which is originally based on an implementation by \citet{dolan2017concurrent}, and turned global allocations into 
local allocations by utilizing the flexibility of the \langkw{region}-construct;
asynchronous computation is based on recursion and tail-call optimization,  
but by omitting the creation of a region around the recursive function, 
we can make local allocations into the region of the enclosing parent function.
Thus, in this case study, we break the association of a region per function body.
\footnote{As mentioned in a previous footnote, 
OxCaml also features special keywords provided whereby the programmer can have finer control over the allocation }
Due to space constraints, we have included the full description of the asynchronous computation case study in \Cref{sec:async-app}.

\section{Conclusion, Related Work and Future Work}
\label{sec:related}
In this paper we introduced Yarrow, an ML-like language 
with algebraic effects and region-based memory management, 
and YL, a separation logic for modular reasoning
about programs written in Yarrow.
We conducted a range of case studies showing 
that programs using one-shot and multi-shot effect can avoid garbage 
collection by using region-based memory management.
We have already discussed related work closely along the way in the paper, 
here we discuss some additional related work and point to future work.

\paragraph{Runtimes for effect handlers} 
A common way of implementing 
effect handlers is using continuation-passing style (CPS) 
\citep{leijen2017type, hillerstrom2020effect}. 
This can be done in intermediate representation languages without 
access to an explicit stack making it more widely applicable to 
different programming languages. In CPS, continuations are allocated as heap 
objects that contain enough information to resume computation. 
This is contrary to the approach by \citet{sivaramakrishnan2021retrofitting} 
that we build upon in this paper where capturing continuations
amounts to saving a pointer to a fiber; the closures we 
associate with continuations are like normal function closure, 
they do not contain all contents of the stack necessary to resume computation as in CPS.
The Flix programming language \citep{flix, flix2} implements effect handlers on top of the Java Virtual Machine (JVM), 
and the language has a region construct. 
The combination of regions and effect handlers is undefined behavior \citep{Magnus}, 
and their region-based memory management has not been treated in any papers.
As mentioned in \Cref{sec:overview}, \citet{Muhcu25} recently proposed a new runtime design for a language with effect handlers (one-shot and multi-shot effects)
and stack references based on the Effekt language \citep{brachthauser2020effects}. 
In this design stack references can be used even after performing multi-shot continuations
by using an \emph{indirection layer}; stack pointers are accessed through indirection pointers. 
A stack reference points to a pointer in the indirection layer which again points to the actual 
stack. This only works for multi-shot effects as long as two continuations never capture the same stack segment, as the case is with
reentrant continuations.
The paper does not come with a program logic or type system for their runtime; 
we suspect that creating a program logic or type system that prohibits 
reentrant continuations would be a significant challenge.

For future work, it is interesting to create a prototype runtime implementation for Yarrow and 
experimentally compare the runtime of programs using effect handlers 
with and without region-based memory management.
The runtime could be complemented with inference of which 
regions have bounded size and can thus be allocated on the call stack
in the style of \citet{birkedal1996region}.
\paragraph{Relational logics and semantic typing} Many Iris-based logics have been created 
to establish relational properties about programs such as \emph{contextual 
equivalence}, see \eg \citep{timany2024logical}. Informally, two programs are contextually equivalent 
if in all contexts they have the same observable behavior.
The main application of 
relational logics and contextual equivalence is to show that an ideal implementation, 
serving as the specification, is equivalent to an efficient algorithm, 
serving as the implementation, e.g. as seen in the works of 
\citet{vindum2021contextual} or \citet{simuliris}.
Recently \citet{de2026relational} made a relational logic to establish
equivalences between programs with effect handlers (they do not prove contextual equivalence).
In the future, we could extend YL to a relational logic.
The extension can be used to establish equivalence between 
complex implementations, optimized w.r.t efficiency by using effect handlers and region-based memory management, 
and counterparts using simpler programming features.
Another direction of future work, is to build a type system for Yarrow 
using \emph{semantic typing} \citep{timany2024logical}. Our logic can serve as a foundation 
in the same way that the original work on effect handlers in separation logic 
by \citet{de2021separation} was used to create a typesystem for 
effect handlers with affine types \citep{Rooij25}, 
and a typesystem for effect handlers with dynamic labels \citep{de2023type}


\bibliography{paper}

\newpage
\appendix
\section{Asynchronous Computation}
\label{sec:async-app}
The asynchronous computation library is implemented using an effect handler, see \Cref{fig:async-comp}.
We focus on the parts of the implementation that we have changed to enable local allocations, 
for a detailed explanation of all the code, we refer to the aforementioned papers.
Asynchronous computation provides two operations: 
\textlang{Async} $\expr$ for asynchronously computing a task represented by the expression $\expr$. 
The \textlang{Async} $\expr$ operation returns a promise $p$ that a user thread can wait for the completion of 
using the \textlang{Await} $p$ operation. The \textlang{Await} $p$ operation returns a value, this is the return 
value of the task that created the promise $p$. 
The implementation in \Cref{fig:async-comp} is based on a function \textlang{fulfill},
this function takes a task $\expr$ and the promise $p$ that the task should fulfill.
Promises are implemented as dynamically allocated references pointing to 
a list of waiting user threads, in our version the references 
are locally allocated. The region that we use to allocate new promises 
in, for every invocation of the \textlang{Async} $\expr$ operation, 
is not the region of the \textlang{fullfill} function that creates promises on line 11;
the \textlang{fullfill} function is recursively called each time an asynchronous computation is made (line 11), 
thus if we want to utilize tail call optimization, we can not rely on the region of \textlang{fullfill}. 
Luckily, the \langkw{region}-construct is flexible in that we can allocate into the region 
of the parent function, \ie \textlang{run}, by omitting creation of a region in \textlang{fullfill}.
Promises are not the only things we allocate into the region of \textlang{run}: 
all the non-empty function closures, on line 7, 11 and 17 use this region too.\footnote{We use the notation "\langkw{fun} $\localVar$ \textlang{arg => e}" when we want to control where a function closure is allocated 
using the locality mode $\localVar$.} 
\begin{figure}[h]
\vspace{-2mm}
\begin{lstlisting}
let new_promise () = ref local (Waiting (list_nil ()))

let next q = if queue_empty q then () else (queue_pop q) ()

let run main = region (
  let q = queue_create () in
  let fulfill = (fun local p e =>
    try (local, once) e () with
    | effect AC request k -> 
      (match request with 
       | Async e -> let p = new_promise () in queue_push q (fun local () => k p); fullfill p e 
       | Await p -> (match !p with Done v -> k v | Waiting ks -> p <- (Waiting (list_cons k ks)); next q))
    | ret x -> 
      (match !p with 
       | Done v -> assert false
       | Waiting ks -> 
         list_iter (fun local k => queue_push q (fun local () => k v)) ks; p <- (Done v); next q)
  ) in
  let p = new_promise () in 
  fullfill p main)
\end{lstlisting}
\caption{Asynchronous Computation.}
\label{fig:async-comp}
\end{figure}

Below, we show the protocol for the asynchronous computation effect $AC$:
\begin{align*}
 \mathit{task} ~ e ~ P ~ \Phi \eqdef ~ & 
    \forall \fdoms_1. ~ P \wand \stackrs ~ (\fdoms_1 \mdoubleplus [(\textlang{AC}, [])]) \wand \mathit{sched} ~ \fdoms_1 \wand \\
        & \hspace{6mm} \ewprec{e ~ ()}{\rowElem{\textlang{AC}}{\onceMode}{~AC}}{\val. ~ \always \Phi ~ \val \ast \exists \fdoms_2, \mathit{sched} ~ \fdoms_2 \ast 
        \stackrs ~ (\fdoms_2 \mdoubleplus [(\textlang{AC}, [])])} \\
  \mathit{AC} \eqdef ~ \hspace{2.5mm} &
   \stackProt{\expr ~ \Phi ~ \fdoms_1 ~ P ~}{\textlang{Async} ~ e, \fdoms_1}{P \ast \mathit{sched} ~ \fdoms_1 \ast 
      \later \mathit{task} ~ e ~ P ~ \Phi}{\onceMode}
    {p ~ \fdoms_2}{p, \fdoms_2}{isPromise ~ p ~ \Phi \ast \mathit{sched} ~ \fdoms_2} \\
  \protComp ~ & \stackProt{p ~ \Phi ~ \fdoms_1}{\textlang{Await} ~ p, \fdoms_1}{isPromise ~ p ~ \Phi \ast \mathit{sched} ~ \fdoms_1}{\onceMode}
    {y ~ \fdoms_2 ~}{y, \fdoms_2}{\always \Phi ~ y \ast \mathit{sched} ~ \fdoms_2}
\end{align*}%
The protocol very closely resembles the one used by \citet{de2021separation}, 
the differences are best seen in the $\mathit{task}$ predicate: 
each new task gets to use the $\stackrs$ resource such that user threads 
can create their own regions, and then we have a predicate $\mathit{sched} ~ \fdoms$. 
The $\mathit{sched} ~ \fdoms$ predicate is abstract to user threads, 
and defined as part of the proof of the effect handler, 
we use it to tie the internal state 
of the effect handler implementation together with allocations in the region 
of the fibers described by fiber domains $\fdoms$.
The postcondition of a task must show ownership of the stack resource $\stackrs$ and the $\mathit{sched}$ 
predicate for some fiber domains $\fdoms_2$ but, importantly, the return value must also 
satisfy the predicate $\Phi$. The predicate $\Phi$ is a persistent predicate that a new task promises to fulfill
when the task is created using the \textlang{Async} operation. 
The \textlang{Async} operation returns a promise $p$ and another persistent predicate $isPromise ~ p ~ \Phi$
stating that $p$ promises to fullfill the proof obligation described by $\Phi$. 
In the specification of the \textlang{Await} $p$ operation, we use $isPromise ~ p ~ \Phi$ as precondition, 
and when the promise is fulfilled, we get to assume $\Phi ~ y$ where $y$ corresponds to the return value 
of the task fulfilling the promise $p$. 
\section{Condition for effect handlers with locally allocated continuation closures}
\label{sec:cont-rule}
\begin{figure}[h!]
\begin{align*}
&\hdlloc ~ \affineVar ~ \prot ~ \Phi_e ~ \var ~ k ~ h ~ y ~ r ~ \rho ~ \mathit{m\rho} ~ \Phi \eqdef{} \\
 &\quad \mathit{m\rho} = \onceMode \rightarrow \onceRow ~ \rho ~ \land\\
 &\quad \condAlways{\mathit{m\rho}}\big( (* ~ Return ~ branch ~ *)\\
 &\qquad\qquad (\forall \val ~ \fdoms. ~ \Phi_e ~ \fdoms ~ \val \wand \stackrs ~ \fdoms ~
    \wand \ewprec{r[\val/y]}{\row}{\Phi}) ~ \land \\
 &\qquad\qquad (* ~ Effect ~ branch ~ *) \\
 &\qquad\qquad (\forall \val ~ k'. ~ (\exists \fdoms ~ \ope ~ \sdoms ~ \sdom ~ \loc ~ \functionid ~ P ~ \Phi_2 ~ \rho' ~ \lctx. ~ 
    \stackrs ~ (\fdoms \mdoubleplus [(\ope, \sdoms \mdoubleplus [\sdom \mdoubleplus [\loc]])]) \ast \loc \stackptr \functionid \asts \\
 &\qquad\qquad\qquad\qquad\quad \prot ~ \val ~ (\fdoms \mdoubleplus [(\ope, \sdoms \mdoubleplus [\sdom])]) ~ 
    (\lambda w_2 ~ \fdoms_2. ~ P ~ \fdoms_2 \wand \ewprec{K[w_2]}{\rho}{\Phi_2}) \asts \\
 &\qquad\qquad\qquad\qquad\quad \condAlways{\affineVar}(\forall w_1 ~ \fdoms_1 ~ \Phi_1. 
    ~ \loc \stackptr \functionid \wand \stackrs ~ \fdoms_1 \wand 
    \hdlloc ~ \affineVar ~ \prot ~ \Phi_e ~ \var ~ k ~ h ~ y ~ r ~ \rho ~ \mathit{m\rho} ~ \Phi_1 \wand \\
 &\qquad\qquad\qquad\qquad\qquad\qquad \later(\loc \stackptr \functionid \wand P ~ \fdoms_1 \wand \ewprec{\lctx[w_1]}{\row'}{\Phi_2}) \wand \\
 &\qquad\qquad\qquad\qquad\qquad\qquad \ewprec{k' ~ w_1}{\row}{\Phi_1})) \wand \\
 &\qquad\qquad\qquad\quad \ewprec{h[k'/k][\val/x]}{\row}{\Phi})\big)
\end{align*}
\caption{Definition of the predicate $\hdlloc$.}
\label{fig:handler-local}
\end{figure}


%

\end{document}